\documentclass[floatfix,twocolumn,prl,amsmath]{revtex4-2}
\usepackage{graphicx}
\usepackage{bm}
\usepackage{amssymb}
\usepackage{dcolumn}
\usepackage{subfigure}
\usepackage{multirow}
\usepackage{mathrsfs}
\DeclareMathAlphabet{\mathscrbf}{OMS}{mdugm}{b}{n}
\usepackage{color}

\begin{document}
\newcommand{\intqspa}{\int\!\!\frac{\rmd^d q}{(2\pi)^d}}
\newcommand{\intqspathr}{\int\!\!\frac{\rmd^3 q}{(2\pi)^3}}
\newcommand{\intqspatwo}{\int\!\!\frac{\rmd^2 q}{(2\pi)^2}}
\newcommand{\intkspatwo}{\int\!\!\frac{\rmd^2 k}{(2\pi)^2}}
\newcommand{\intkspa}{\int\!\!\frac{\rmd^d k}{(2\pi)^d}}
\newcommand{\intkspapri}{\int\!\!\frac{\rmd^d k'}{(2\pi)^d}}
\newcommand{\vn}[1]{{\boldsymbol{#1}}}
\newcommand{\vht}[1]{{\boldsymbol{#1}}}
\newcommand{\matn}[1]{{\bf{#1}}}
\newcommand{\matnht}[1]{{\boldsymbol{#1}}}
\newcommand{\bege}{\begin{equation}}
\newcommand{\gretke}{G_{\vn{k} }^{\rm R}(\mathcal{E})}
\newcommand{\gret}{G^{\rm R}}
\newcommand{\gadv}{G^{\rm A}}
\newcommand{\gmat}{G^{\rm M}}
\newcommand{\gles}{G^{<}}
\newcommand{\ghat}{\hat{G}}
\newcommand{\sigmahat}{\hat{\Sigma}}
\newcommand{\glesone}{G^{<,{\rm I}}}
\newcommand{\glestwo}{G^{<,{\rm II}}}
\newcommand{\gspec}{G^{\rm S}}
\newcommand{\glesthree}{G^{<,{\rm III}}}
\newcommand{\magdir}{\hat{\vn{n}}}
\newcommand{\sigmaret}{\Sigma^{\rm R}}
\newcommand{\sigmales}{\Sigma^{<}}
\newcommand{\sigmalesone}{\Sigma^{<,{\rm I}}}
\newcommand{\sigmalestwo}{\Sigma^{<,{\rm II}}}
\newcommand{\sigmalesthree}{\Sigma^{<,{\rm III}}}
\newcommand{\sigmaadv}{\Sigma^{A}}
\newcommand{\ee}{\end{equation}}
\newcommand{\bal}{\begin{aligned}}
\newcommand{\defbar}{\overline}
\newcommand{\SM}{\scriptstyle}
\newcommand{\rmd}{{\rm d}}
\newcommand{\rme}{{\rm e}}
\newcommand{\eal}{\end{aligned}}
\newcommand{\crea}[1]{{c_{#1}^{\dagger}}}
\newcommand{\annihi}[1]{{c_{#1}^{\phantom{\dagger}}}}
\newcommand{\udot}{\overset{.}{u}}
\newcommand{\exponential}[1]{{\exp(#1)}}
\newcommand{\phandot}[1]{\overset{\phantom{.}}{#1}}
\newcommand{\phandag}{\phantom{\dagger}}
\newcommand{\Trace}{\text{Tr}}
\setcounter{secnumdepth}{2}
\title{Construction of the spectral function from noncommuting spectral moment matrices}
\author{Frank Freimuth$^{1,2}$}
\email[Corresp.~author:~]{f.freimuth@fz-juelich.de}
\author{Stefan Bl\"ugel$^{1}$}
\author{Yuriy Mokrousov$^{1,2}$}
\affiliation{$^1$Peter Gr\"unberg Institut and Institute for Advanced Simulation,
Forschungszentrum J\"ulich and JARA, 52425 J\"ulich, Germany}
\affiliation{$^2$ Institute of Physics, Johannes Gutenberg University Mainz, 55099 Mainz, Germany
}
\begin{abstract}
  The LDA+$U$ method is widely used to study the properties of realistic solids with
  strong electron correlations. One of its main shortcomings is that 
  it does not provide direct access to the temperature dependence of material properties
  such as the magnetization, the magnetic anisotropy energy, the Dzyaloshinskii-Moriya interaction,
  the anomalous Hall conductivity, and the spin-orbit torque.
  While the method of spectral moments allows us in principle to compute these quantities directly at finite
  temperatures,
  the standard two-pole approximation can be applied only to Hamiltonians that are
  effectively of single-band type.  
  We do a first step to explore if the method of spectral moments may replace the
  LDA+$U$ method in first-principles calculations of correlated solids with many bands in cases where the
  direct assessment of the temperature dependence of
  equilibrium and response functions is desired:
  The spectral moments of many-band Hamiltonians of correlated electrons do not commute and therefore
  they do not possess a system of common eigenvectors.
  We show that nevertheless the spectral function may be constructed from the spectral moments
  by solving a system of coupled non-linear equations.
  Additionally, we show how to compute the anomalous Hall conductivity of correlated
  electrons from this spectral function. We demonstrate the method for the Hubbard-Rashba model,
  where the standard two-pole approximation cannot be applied, because spin-orbit interaction (SOI)
  couples the spin-up and the spin-down bands.
  In the quest for new quantum states that arise from the combination of
  SOI and correlation effects, the Hartree-Fock approximation is frequently used
  to obtain a first approximation for the phase diagram. We propose that using the many-band
  generalization of the selfconsistent moment method instead of Hartree-Fock in such
  exploratory model calculations
  may improve the accuracy significantly, while keeping the computational burden low.
\end{abstract}
\maketitle
\section{Introduction}
The LDA+$U$ approach is a very popular method to compute the electronic structure of
realistic solids with strong electron
correlations~\cite{review_ldau,PhysRevB.44.943,PhysRevB.48.16929,PhysRevB.50.16861}.
While it is less accurate than dynamical mean field theory (DMFT)~\cite{PhysRevLett.87.067205}
it is computationally faster.

Recently, the demand to compute spintronic material properties  
such as the Dzyaloshinskii-Moriya interaction~\cite{doi:10.1063/1.5038353},
the spin-orbit torque~\cite{PhysRevB.89.174424,angular_and_temperature_dependence_Ta_CoFeB_MgO,magnonicsot,rmp_sot}
and the Hall coefficient~\cite{dc_hall_Hubbard_Wang2020,PhysRevResearch.3.033033,PhysRevB.99.115115,PhysRevLett.121.066601}
at finite temperatures has increased.
While LDA+DMFT provides direct access to the temperature dependence
of equilibrium and response property tensors, the LDA+$U$ scheme is limited to zero temperature unless
it is extended by subsequent calculations of the effects of finite temperature e.g.\ by
Monte-Carlo simulations, by classical atomistic spin models, 
or by Green's function theory~\cite{PhysRevB.96.094436,CALLEN19661271,PhysRev.130.890}.
This poses the question of whether extending LDA+$U$ by additional higher-order correlation
functions beyond the mean-field level will allow us to compute these material
properties directly at finite temperature with a computational cost comparable to LDA+$U$ calculations,
i.e., a computational cost much smaller than the one of LDA+DMFT.
This question is particularly important in view of response tensors such as the
anomalous Hall effect (AHE)~\cite{PhysRevLett.92.037204,RevModPhys.82.1539,PhysRevB.74.195118} or the spin-orbit torque~\cite{ibcsoit},
which often require a much finer
$k$-point sampling to achieve convergence than the calculation of total energies and magnetic moments
does, e.g.\ due to the spiky Berry
curvature of the AHE.

The self-consistent spectral moment method explored in the eighties on the basis of a single band Hubbard model  is an alternative approach  able to predict realistic results for the Curie temperatures in Fe and Ni~\cite{bcc_iron_Nolting1995,nickel_PhysRevB.40.5015,nickel_Borgiel1990}
and it is computationally much less demanding than LDA+DMFT. 
Similarly to LDA+$U$ and LDA+DMFT, it improves
LDA by adding a Hubbard-type interaction to it. 
While the Kubo-Bastin equation for the
electrical conductivity~\cite{crepieux_bruno_ahe,bastin_conductivity} does not
hold for correlated electrons, the method of spectral moments can also be used to compute the
response functions avoiding the independent particle approximation
of the Kubo-Bastin equation~\cite{hubbard_model_electrical_conductivity}.
In this paper we embark on extending the self-consistent spectral moment
approach to a new practical computational method that can be combined
with density functional theory to study the finite temperature
properties of realistic magnetic materials and spintronic
modules. For this we need to go beyond the single band approximation
of the original method. The success of our new approach
is explicitly shown for the relativistic Rashba Hubbard model.
Avoiding the solution of the quantum impurity problem inherent to the
LDA-DMFT method, our approach to the finite temperature problem
is expected to be computationally much faster.

We define the $n$-th spectral moment matrix
at $k$-point $\vn{k}$ by~\cite{bcc_iron_Nolting1995,nickel_PhysRevB.40.5015,PhysRevB.38.2608}
\bege\label{eq_specmoms_eneint}
\vn{M}^{(n)}_{\vn{k}}=\frac{1}{\hbar}\int_{-\infty}^{\infty} \vn{S}_{\vn{k}}(E) E^{n}  dE,
\ee
where $\vn{S}_{\vn{k}}(E)$ is the spectral density matrix at energy $E$.
Since the spectral moments may alternatively be expressed in terms
of thermal averages of anticommutators such as~\cite{bcc_iron_Nolting1995,nickel_PhysRevB.40.5015,PhysRevB.38.2608}
\bege\label{eq_specmom_expectvalue}
\tilde{M}^{(1)}_{\vn{k}\alpha\beta}=\frac{1}{N}
\sum_{ij}e^{i\vn{k}\cdot
  (\vn{R}_{i}-\vn{R}_{j})}
\langle[[c_{i\alpha},\mathcal{H}]_{-},c^{\dagger}_{j\beta}]_{+} \rangle,
\ee
where $c^{\dagger}_{j\beta}$ creates an electron in state $\beta$ at lattice
site $j$ (position $\vn{R}_{j}$), $c_{i\alpha}$ annihilates an electron in
state $\alpha$ at lattice
site $i$  (position $\vn{R}_{i}$), $\mathcal{H}=H-\mu \hat{N}$,  
$H$ is the Hamiltonian of the system, $\hat{N}$ is the number operator,
$\mu$ is the chemical potential, $[\dots]_{-}$ denotes the commutator,
and $[\dots]_{+}$ is the anticommutator, a closed system of equations
for $\vn{S}_{\vn{k}}(E)$
may be obtained by requiring that $\vn{M}^{(n)}_{\vn{k}}=\tilde{\vn{M}}^{(n)}_{\vn{k}}$.
Usually, the first four moments are considered, i.e., n=0, 1, 2, 3.
In Eq.~\eqref{eq_specmom_expectvalue} we give only the
anticommutator for the first moment. The $n$-th moment contains
$n$ commutators, which are generated by applying the equation of motion
$n$ times. Explicit expressions are given in App.~\ref{sec_appendix_hubbard}.

In the single-band Hubbard model the electron can only be in the
state of spin up ($\uparrow$) or spin down ($\downarrow$), i.e.,
$\alpha=\uparrow,\downarrow$, and one may choose the spectral
function and the spectral moments to be diagonal in the spin indices,
i.e., $M^{(n)}_{\vn{k}\alpha\beta}=M^{(n)}_{\vn{k}\alpha}\delta_{\alpha\beta}$
and $S_{\vn{k}\alpha\beta}=S_{\vn{k}\alpha}\delta_{\alpha\beta}$.
Considering the first 4 moments -- n=0, 1, 2, 3 -- one obtains
4 equations for $\alpha=\uparrow$
and 4 equations for $\alpha=\downarrow$. These 4 equations may be used
to determine the 4 unknown coefficients
$a_{\vn{k}\alpha 1}$, $a_{\vn{k}\alpha 2}$, $E_{\vn{k}\alpha 1}$, and $E_{\vn{k}\alpha 2}$
in the two-pole
approximation~\cite{PhysRevB.38.2608} 
\bege\label{eq_twopole_twoband}
\frac{S_{\vn{k}\alpha}(E-\mu)}{\hbar}=a_{\vn{k}\alpha 1}\delta(E-E_{\vn{k}\alpha 1})
+a_{\vn{k}\alpha 2}\delta(E-E_{\vn{k}\alpha 2})
\ee
of the spectral density.
Here, $E_{\vn{k}\alpha 1}$ and $a_{\vn{k}\alpha 1}$ are the band energy and the
spectral weight of the lower Hubbard band, while
$E_{\vn{k}\alpha 2}$ and $a_{\vn{k}\alpha 2}$ are those of the upper Hubbard band.
The alternative expression of the third moment $M^{(3)}_{\vn{k}\uparrow}$
in terms of the thermal average of an anticommutator contains higher-order correlations
such
as $\langle c^{\dagger}_{i\uparrow}c^{\dagger}_{i\downarrow}c_{j\downarrow} c_{j\uparrow}\rangle$,
which lead to a much more realistic description of the effect of finite temperature
than the mean field level can provide.

In order to replace LDA+$U$ by the method of spectral moments
one might consider to map the electronic structure of a standard
LDA calculation
first onto a set of maximally localized Wannier
functions (MLWFs)~\cite{wannier90communitycode}.
$c_{i\alpha}$ may then be considered to be the annihilation operator
of an electron in the MLWF state $|\vn{R}_{i}\alpha\rangle$.
Note that MLWFs are routinely used for the Wannier interpolation
of response functions such as the AHE~\cite{PhysRevB.74.195118} and the
spin-orbit torque~\cite{ibcsoit}. When one wishes to compute such response
functions, the mapping of the electronic structure onto MLWFs
is therefore often not an additional complication, because this step
will have to be done in any case.
Some implementations of LDA+$U$ already use Wannier functions as
basis set~\cite{ldau_wannier_Anisimov_2007}.

The question now arises of how to modify
Eq.~\eqref{eq_twopole_twoband}
when we have $N_{\rm W}$ states, i.e., $\alpha=1,\dots, N_{\rm W}$ instead
of only the two states $\alpha=\uparrow,\downarrow$ in the single-band
Hubbard model.
An obvious generalization of Eq.~\eqref{eq_twopole_twoband}
is given by
\bege\label{eq_spectral_matrix_general}
\frac{S_{\vn{k}\alpha\beta}(E-\mu)}{\hbar}=\sum_{p=1}^{2}\sum_{\gamma=1}^{N_{\rm W}}
a_{\vn{k}\gamma p}
\mathcal{V}_{\vn{k}\alpha\gamma p}\mathcal{V}^{*}_{\vn{k}\beta\gamma p} \delta(E-E_{\vn{k}\gamma p}),
\ee
where $p=1$ denotes the lower Hubbard band, while $p=2$ denotes the upper one,
and $\mathcal{V}_{\vn{k}\alpha\gamma p}$ is the component $\alpha$ of a normalized vector
associated with the band described by the indices $\gamma$ and $p$ with
the band energy $E_{\vn{k}\gamma p}$. This approximation for $S_{\vn{k}\alpha\beta}(E)$
is hermitean, i.e., $S_{\vn{k}\alpha\beta}(E)=[S_{\vn{k}\beta\alpha}(E)]^{*}$. 
The unknown parameters $E_{\vn{k}\gamma p}$, $a_{\vn{k}\gamma p}$,  and $\mathcal{V}_{\vn{k}\alpha\gamma p}$
need to be determined by equating Eq.~\eqref{eq_specmoms_eneint}
and Eq.~\eqref{eq_specmom_expectvalue}.
To demonstrate how this may be achieved is the central goal of this paper.

The difficulty to determine the unknown parameters
in Eq.~\eqref{eq_spectral_matrix_general}
may be illustrated by comparing it to the spectral function
of a closed system with $N_{\rm W}$ independent electrons,
which is given by
\bege\label{eq_spectral_matrix_nonint}
\bar{S}_{\vn{k}\alpha\beta}(E-\mu)=\hbar\sum_{\gamma=1}^{N_{\rm W}}
\mathcal{U}_{\vn{k}\alpha\gamma}\mathcal{U}^{*}_{\vn{k}\beta\gamma} \delta(E-E_{\vn{k}\gamma}),
\ee
where the unitary $N_{\rm W}\times N_{\rm W}$
matrices $\vn{\mathcal{U}}_{\vn{k}}$ describe the
transformation between the basis functions $|\phi_{\vn{k}\beta}\rangle$
and the eigenfunctions $|\psi_{\vn{k}\alpha}\rangle$
\bege\label{eq_unitary_trafo}
|\psi_{\vn{k}\alpha}\rangle=\sum_{\beta} \mathcal{U}_{\vn{k}\beta\alpha}  |\phi_{\vn{k}\beta}\rangle,
\ee
which are eigenfunctions with energy $E_{\vn{k}\alpha}$.
Clearly, the $N_{\rm W}$ columns of the matrix $\vn{\mathcal{U}}_{\vn{k}}$
are the $N_{\rm W}$ eigenvectors of the moments $\vn{M}_{\vn{k}}^{(n)}$
of $\vn{\bar{S}}_{\vn{k}}(E)$ with eigenvalues $(E_{\vn{k}\alpha}-\mu)^{n}$.
In contrast, the unknown vectors $\mathcal{V}_{\vn{k}\alpha\gamma p}$
are generally not the eigenvectors of the spectral moments $\vn{M}_{\vn{k}}^{(1)}$,
$\vn{M}_{\vn{k}}^{(2)}$, and $\vn{M}_{\vn{k}}^{(3)}$, because these spectral moments
do not commute for correlated electrons and therefore they do not possess a common
system of eigenvectors.

Another argument to formulate this difficulty is as follows:
The band $\gamma$ splits into the lower Hubbard band and the upper Hubbard band and
generally $\mathcal{V}_{\vn{k}\alpha\gamma 1}\ne \mathcal{V}_{\vn{k}\alpha\gamma 2}$.
Consequently, we need to determine $2N_{\rm W}$ state vectors, but each spectral
moment matrix $\vn{M}_{\vn{k}}^{(n)}$ has only $N_{\rm W}$ eigenvectors. Thus, the
eigenvectors of the spectral matrices are not eigenstates of the Hamiltonian
in the case of correlated electrons, which is a major difference when compared with
closed systems of uncorrelated electrons.

The interplay of spin-orbit interaction (SOI) with electron correlations
may lead to new and exotic quantum
phases~\cite{corr_plus_soi_balents,square_lattice_iridates}.
In order to obtain a first approximation for the phase diagram the
Hartree-Fock approximation is frequently used in
order to explore new quantum states that arise from the combination
of SOI and correlation
effects~\cite{PhysRevB.85.045124,PhysRevB.87.155101,metal_insulator_rashba_hubbard}.
Since the treatment of the higher-order
correlation $\langle c^{\dagger}_{i\alpha}c^{\dagger}_{j\beta}c_{l\gamma} c_{m\delta}\rangle$
within the selfconsistent moment method goes beyond the Hartree-Fock level,
one may expect that this method may improve the accuracy of the phase diagram
significantly, while keeping the computational effort low.
Broken space inversion symmetry gives rise to 
Rashba-type SOI, which promotes many important
effects in spintronics~\cite{Bercioux_2015,rashba_review}.
The Hubbard-Rashba
model~\cite{rashbahub_PhysRevB.88.045102,metal_insulator_rashba_hubbard}
combines 
the Rashba-type SOI with electron correlations.
Since the Rashba-type SOI couples the spin-up and spin-down
bands, the two-pole approximation Eq.~\eqref{eq_twopole_twoband}
cannot be applied to the Hubbard-Rashba model.
Therefore, we take the Hubbard-Rashba model as an example
to demonstrate our method to construct the spectral function.

The rest of this paper is organized as follows:
In Sec.~\ref{sec_ground_state}
we describe how to determine the unknown
parameters $E_{\vn{k}\gamma p}$, $a_{\vn{k}\gamma p}$,  and $\mathcal{V}_{\vn{k}\alpha\gamma p}$
in Eq.~\eqref{eq_spectral_matrix_general}
in order to determine the ground state and the bandstructure of the
correlated system.
In Sec.~\ref{sec_corr_fun} we discuss how to compute
correlation functions of the
type $\langle c^{\dagger}_{i\alpha}c^{\dagger}_{j\beta}c_{l\gamma} c_{m\delta}\rangle$
using the method of spectral moments. These correlation functions
are needed to compute the third moments $\tilde{M}^{(3)}_{\vn{k}\alpha\beta}$.
In Sec.~\ref{sec_indep_electrons} we discuss how to obtain
response functions based on the method of spectral moments
for systems of independent electrons. This section provides
useful guidelines to determine the response functions for
correlated electrons in Sec.~\ref{sec_resp_corr}.
In Sec.~\ref{sec_rashba_hubbard}
we demonstrate the method for the Hubbard-Rashba model.
Sec.~\ref{sec_outlook} provides an outlook of how this approach
may be combined with DFT.
This paper ends with a summary in Sec.~\ref{sec_summary}.

\section{Ground state and bandstructure}
\label{sec_ground_state}
The moments $\tilde{\vn{M}}^{(n)}_{\vn{k}}$ (Eq.~\eqref{eq_specmom_expectvalue})
are hermitean $N_{\rm W}\times N_{\rm W}$ matrices.
A hermitean $N_{\rm W}\times N_{\rm W}$ matrix
is fully defined by $N_{\rm W}^{2}$ real-valued parameters, because its diagonal
is real-valued and the upper triagonal is the complex-conjugate of the lower triagonal.
Consequently, we may map each moment $\tilde{\vn{M}}^{(n)}_{\vn{k}}$
onto an $N_{\rm W}^{2}$-dimensional real-valued vector
$\vn{\mathcal{M}}^{(n)}_{\vn{k}}$.
To be concrete, we fill the first $N_{\rm W}(N_{\rm W}+1)/2$ components
of the vector $\vn{\mathcal{M}}^{(n)}_{\vn{k}}$ with the real parts of the
elements from the upper triagonal of $\tilde{\vn{M}}^{(n)}_{\vn{k}}$, and
the remaining $N_{\rm W}(N_{\rm W}-1)/2$ components we fill with the
imaginary parts of the
elements from the upper triagonal.
We define the $N_{\rm W}^{2}\times 4$ matrix
$\vn{\mathcal{M}}_{\vn{k}}$ by
\bege
\vn{\mathcal{M}}_{\vn{k}}=[\vn{\mathcal{M}}^{(0)}_{\vn{k}},\vn{\mathcal{M}}^{(1)}_{\vn{k}},\vn{\mathcal{M}}^{(2)}_{\vn{k}},\vn{\mathcal{M}}^{(3)}_{\vn{k}}].
\ee

Inserting the approximation Eq.~\eqref{eq_spectral_matrix_general}
into Eq.~\eqref{eq_specmoms_eneint} yields
\bege\label{eq_spectral_matrix_general_eneint}
M^{(n)}_{\vn{k}\alpha\beta}=
\sum_{p=1}^{2}\sum_{\gamma=1}^{N_{\rm W}}
a_{\vn{k}\gamma p}
\mathcal{W}_{\vn{k}\alpha\beta\gamma p}
(E_{\vn{k}\gamma p}-\mu)^n,
\ee
where we
defined $\mathcal{W}_{\vn{k}\alpha\beta\gamma p}=\mathcal{V}_{\vn{k}\alpha\gamma p}\mathcal{V}^{*}_{\vn{k}\beta\gamma p}$.
We may consider $\mathcal{W}_{\vn{k}\alpha\beta\gamma p}$ as the row-$\alpha$ column-$\beta$ element of
a hermitean matrix $\vn{\mathcal{W}}_{\vn{k}\gamma p}$.
Since $\gamma=1,\dots,N_{\rm W}$ and $p=1,2$, there are $2N_{\rm W}$ such matrices at a given $k$-point $\vn{k}$.
As the hermitean $N_{\rm W}\times N_{\rm W}$ matrix $\vn{\mathcal{W}}_{\vn{k}\gamma p}$  is
equivalent to a $N_{\rm W}^2$-dimensional real-valued vector $\tilde{\vn{\mathcal{W}}}_{\vn{k}\gamma p}$,
we define the $N_{\rm W}^2 \times 2N_{\rm W}$
matrix $\vn{\mathcal{W}}_{\vn{k}}=[\tilde{\vn{\mathcal{W}}}_{\vn{k}11}\dots \tilde{\vn{\mathcal{W}}}_{\vn{k}N_{\rm W}2}]$.
Additionally, we construct the $2N_{\rm W}\times 4$ matrix $\vn{\mathcal{A}}_{\vn{k}}$ by setting the element
$\mathcal{A}_{\vn{k}\gamma p m}$ in row $(\gamma, p)$ and
column $m$ to $a_{\vn{k}\gamma p}(E_{\vn{k}\gamma p}-\mu)^{m-1}$.
The requirements $\vn{M}^{(n)}_{\vn{k}}=\tilde{\vn{M}}^{(n)}_{\vn{k}}$ with n=0, 1, 2, 3
can now be formulated in compact form by
\bege\label{eq_compact_m_eq_m}
\vn{\mathcal{W}}_{\vn{k}}\vn{\mathcal{A}}_{\vn{k}}=\vn{\mathcal{M}}_{\vn{k}}.
\ee

The moments computed from the thermal averages of anticommutator expressions, e.g.\
Eq.~\eqref{eq_specmom_expectvalue},
are stored in the matrix $\vn{\mathcal{M}}_{\vn{k}}$. In the step of solving
Eq.~\eqref{eq_compact_m_eq_m} they are considered as fixed given input.
The unknown band energies and spectral weights are contained
in the matrix $\vn{\mathcal{A}}_{\vn{k}}$.
The matrix $\vn{\mathcal{W}}_{\vn{k}}$ is constructed from the unknown state
vectors $\vn{\mathcal{V}}_{\vn{k}\gamma p}$.

Eq.~\eqref{eq_compact_m_eq_m} defines 
$4N_{\rm W}^2$ nonlinear equations, because this is the number of matrix elements
in $\vn{\mathcal{M}}_{\vn{k}}$.
Each vector $\vn{\mathcal{V}}_{\vn{k}\gamma p}$ has $N_{\rm W}$ components and there are
$2N_{\rm W}$ such vectors. Since $\vn{\mathcal{V}}_{\vn{k}\gamma p}$ is required to be
a normalized vector and since the
gauge-transformation $\vn{\mathcal{V}}_{\vn{k}\gamma p}\rightarrow e^{i\Phi}\vn{\mathcal{V}}_{\vn{k}\gamma p} $
has no effect on $\mathcal{W}_{\vn{k}\alpha\beta\gamma p}=\mathcal{V}_{\vn{k}\alpha\gamma p}\mathcal{V}^{*}_{\vn{k}\beta\gamma p}$,
every $\vn{\mathcal{V}}_{\vn{k}\gamma p}$ is determined by $2(N_{\rm W}-1)$ real-valued unknowns, i.e.,
$4(N_{\rm W}^2-N_{\rm W})$ unknown coefficients need to be determined to define all vectors $\vn{\mathcal{V}}_{\vn{k}\gamma p}$.
Additionally, we need to determine the $2N_{\rm W}$ unknown energies $E_{\vn{k}\gamma p}$
and the $2N_{\rm W}$ unknown spectral weights $a_{\vn{k}\gamma p}$. Thus, Eq.~\eqref{eq_compact_m_eq_m}
defines a system of $4N_{\rm W}^2$ nonlinear equations for $4N_{\rm W}^2$ unknown parameters.
Since the number of unknowns matches the number of available equations, our approximation
Eq.~\eqref{eq_spectral_matrix_general}
for the spectral function is justified.
This proves that the spectral function may be constructed from the
spectral moments even in the correlated many-band case, which is a central
result of this paper.

In order to solve Eq.~\eqref{eq_compact_m_eq_m} numerically one may use
the standard equation solvers available in many mathematical libraries.
These equation solvers typically require the user to formulate the problem in the form
\bege
f_{m}(\vn{x})=0,
\ee
where the components of the $4N_{\rm W}^2$-dimensional
vector $\vn{x}$ are the unknown coefficients and
the $4N_{\rm W}^2$ functions $f_{m}(\vn{x})$ may be obtained by
rewriting Eq.~\eqref{eq_compact_m_eq_m} as
\bege
\vn{\mathcal{F}}_{\vn{k}}=\vn{\mathcal{W}}_{\vn{k}}\vn{\mathcal{A}}_{\vn{k}}-\vn{\mathcal{M}}_{\vn{k}}
\ee
and setting the $m$-th function $f_{m}(\vn{x})$ to the $m$-th entry of the
matrix $\vn{\mathcal{F}}_{\vn{k}}$, which has $4N_{\rm W}^2$ entries in total.
From the equations above it is clear that it is straightforward to obtain
expressions for 
the derivatives $df_{m}(\vn{x})/dx_{j}$. Thus, one may also provide the Jacobian
to the equation solver.

The calculation of the moments $\tilde{M}^{(n)}_{\vn{k}\alpha\beta}$
requires the thermal averages $\langle c^{\dagger}_{i\alpha} c_{j\beta}\rangle$
as input. These thermal averages can be computed easily from the spectral
function $\vn{S}_{\vn{k}}(E)$, which we discuss in the next section.
However, since the spectral function itself needs to be computed, a self-consistency
scheme is required:
A starting guess for $\langle c^{\dagger}_{i\alpha} c_{j\beta}\rangle$ needs to be chosen.
Using this starting guess, one may then evaluate the spectral moments
$\tilde{M}^{(n)}_{\vn{k}\alpha\beta}$
for $n=0,1,2,3$. Examples for the expressions of $\tilde{M}^{(n)}_{\vn{k}\alpha\beta}$
are given in Appendix~\ref{sec_appendix_hubbard} for the Hubbard-Rashba model. 
The moments $\tilde{M}^{(n)}_{\vn{k}\alpha\beta}$ determine the right-hand side
of Eq.~\eqref{eq_compact_m_eq_m} completely.
Next, one solves Eq.~\eqref{eq_compact_m_eq_m} for the unknown
coefficients $E_{\vn{k}\gamma p}$, $a_{\vn{k}\gamma p}$, and $\vn{\mathcal{V}}_{\vn{k}\gamma p}$,
which determine the left-hand side of Eq.~\eqref{eq_compact_m_eq_m}.
Using these, one evaluates the spectral function
according to Eq.~\eqref{eq_spectral_matrix_general}.
Now, one may compute new averages $\langle c^{\dagger}_{i\alpha} c_{j\beta}\rangle$
employing this spectral function.
This completes the first iteration of the self-consistency loop.
At the beginning of the next iteration,
one uses the new averages $\langle c^{\dagger}_{i\alpha} c_{j\beta}\rangle$
to compute the moments $\tilde{M}^{(n)}_{\vn{k}\alpha\beta}$.
The procedure is repeated until self-consistency is reached, i.e., it
is repeated until the averages $\langle c^{\dagger}_{i\alpha} c_{j\beta}\rangle$ used as input to evaluate
the spectral moments $\tilde{M}^{(n)}_{\vn{k}\alpha\beta}$
agree with the output averages $\langle c^{\dagger}_{i\alpha} c_{j\beta}\rangle$.

For the calculation of the moments $\tilde{M}^{(3)}_{\vn{k}\alpha\beta}$ we may need
in addition correlation functions of the
type $\langle c^{\dagger}_{i\alpha}c^{\dagger}_{j\beta}c_{l\gamma} c_{m\delta}\rangle$.
Their computation is discussed in the next section and it
may require correlation
functions $\langle c^{\dagger}_{i'\alpha'}c^{\dagger}_{j'\beta'}c_{l'\gamma'} c_{m'\delta'}\rangle$ as input.
Therefore, in order to
compute $\langle c^{\dagger}_{i\alpha}c^{\dagger}_{j\beta}c_{l\gamma} c_{m\delta}\rangle$
we use a self-consistent procedure as well
and we include the correlation functions
$\langle c^{\dagger}_{i\alpha}c^{\dagger}_{j\beta}c_{l\gamma} c_{m\delta}\rangle$
into the selfconsistency scheme: Not
only $\langle c^{\dagger}_{i\alpha} c_{j\beta}\rangle$  but also
the correlation
functions $\langle c^{\dagger}_{i\alpha}c^{\dagger}_{j\beta}c_{l\gamma} c_{m\delta}\rangle$
are computed self-consistently, i.e., the calculation is considered converged when
their computed output agrees to their input.

The question poses itself if one may interpret Eq.~\eqref{eq_compact_m_eq_m} 
as a generalization of the well-known eigenvalue problem of matrix diagonalization,
which needs to be solved in the case of closed systems of independent particles.
Therefore, let us recall this non-interacting case.
Inserting Eq.~\eqref{eq_spectral_matrix_nonint}
into Eq.~\eqref{eq_specmoms_eneint}
yields
\bege\label{eq_eigenvalue_probl}
M_{\vn{k}\alpha\beta}^{(n)}=\sum_{\gamma=1}^{N_{\rm W}}
\mathcal{U}_{\vn{k}\alpha\gamma}\mathcal{U}^{*}_{\vn{k}\beta\gamma}[E_{\vn{k}\gamma}-\mu]^{n}.
\ee
Therefore, one may pick any value $n>0$ and obtain the eigenvalues from the
spectral moment matrix $\vn{M}^{(n)}_{\vn{k}}$.
Since the eigenvalue problem of matrix diagonalization is an important problem
in linear algebra, one usually uses linear algebra algorithms to
diagonalize $\vn{M}^{(n)}_{\vn{k}}$ and to 
find thereby
the eigenvalues $[E_{\vn{k}\gamma}-\mu]^{n}$
and the eigenvectors, which are the columns of the unitary matrix $\vn{\mathcal{U}}_{\vn{k}}$.

However, one may easily cast the problem Eq.~\eqref{eq_eigenvalue_probl}
into the form of Eq.~\eqref{eq_compact_m_eq_m}:
We may
consider $\bar{\mathcal{W}}_{\vn{k}\alpha\beta\gamma}=\mathcal{U}_{\vn{k}\alpha\gamma}\mathcal{U}^{*}_{\vn{k}\beta\gamma}$
as the row-$\alpha$ column-$\beta$ element of
a hermitean matrix $\bar{\vn{\mathcal{W}}}_{\vn{k}\gamma }$,
rewrite it as a $N_{W}^2$-dimensional vector and collect all these $N_{\rm W}$
vectors ($\gamma=1\dots N_{\rm W}$) in
the $N_{\rm W}^2\times N_{\rm W}$ matrix $\bar{\vn{\mathcal{W}}}_{\vn{k}}$.
If we map the moment $\vn{M}^{(n)}_{\vn{k}}$ onto an $N_{W}^2$-dimensional
vector $\vn{\mathcal{\vn{M}}}^{(n)}_{\vn{k}}$,
we may write Eq.~\eqref{eq_eigenvalue_probl}
in the form of Eq.~\eqref{eq_compact_m_eq_m}:
\bege\label{eq_revisit_eigenvalueprobl}
\bar{\vn{\mathcal{\vn{W}}}}_{\vn{k}}\vn{\mathcal{A}}^{(n)}_{\vn{k}}=\vn{\mathcal{\vn{M}}}^{(n)}_{\vn{k}},
\ee
where the $\gamma$-th entry in the vector $\vn{\mathcal{A}}^{(n)}_{\vn{k}}$
is the eigenvalue $[E_{\vn{k}\gamma}-\mu]^{n}$.

The Eq.~\eqref{eq_revisit_eigenvalueprobl} defines $N_{\rm W}^2$ equations.
The unitary matrix $\vn{\mathcal{U}}_{\vn{k}}$ contains $N_{\rm W}$ normalized vectors. Since their phase
does not matter, only $2N_{\rm W}^2-2N_{\rm W}$ real parameters are needed to determine these vectors.
However, the columns of the matrix $\vn{\mathcal{U}}_{\vn{k}}$ are mutually orthogonal.
This reduces the number of independent real parameters needed to determine $\vn{\mathcal{U}}_{\vn{k}}$
by $N_{\rm W}^2-N_{\rm W}$. Thus, we need only $N_{\rm W}^2-N_{\rm W}$ parameters to
determine $\bar{\vn{\mathcal{\vn{W}}}}_{\vn{k}}$. Since we need to find
the $N_{\rm W}$ eigenvalues  $[E_{\vn{k}\gamma}-\mu]^{n}$
as well, $N_{\rm W}^2$ parameters need to be found in total, which matches the number
of available equations. Therefore, one may argue that Eq.~\eqref{eq_compact_m_eq_m}
is a generalization of the standard eigenvalue problem, which needs to be used
when the bands of correlated electron systems split into lower and upper Hubbard bands,
while it is equivalent to the standard eigenvalue problem when the independent
particle approximation is used, i.e., when the upper Hubbard bands
are not considered.

\section{Correlation functions}
\label{sec_corr_fun}
In Appendix~\ref{sec_appendix_hubbard}
we give explicit expressions for the first
4 moments of the single-particle spectral function
for the example of the Hubbard-Rashba model.
In these expressions correlation functions of the
types
$\langle c^{\dagger}_{i\alpha} c_{j\beta} \rangle$
and
$\langle c^{\dagger}_{i\alpha}  c^{\dagger}_{j\beta} c_{l\gamma} c_{m\delta} \rangle$
occur.
These correlation functions may be computed
from the corresponding anticommutator spectral functions
using the spectral theorem~\cite{book_Nolting}:
\bege\label{eq_spec_theo_single_part}
\langle c^{\dagger}_{\vn{k}\alpha} c_{\vn{k}\beta}\rangle=\frac{1}{\hbar}
\int_{-\infty}^{\infty}d\, E
f(E)
S_{\vn{k}\beta\alpha}(E-\mu),
\ee
and
\bege\label{eq_corr4_specfun}
\langle c^{\dagger}_{\vn{k}\alpha}  c^{\dagger}_{j\beta} c_{l\gamma} c_{m\delta} \rangle=
\frac{1}{\hbar}
\int_{-\infty}^{\infty}d\, E
f(E)
S_{l\gamma m\delta j\beta \vn{k}\alpha}(E-\mu),
\ee
where $f(E)$ is the Fermi-Dirac distribution function.

In the case of $\langle c^{\dagger}_{\vn{k}\alpha} c_{\vn{k}\beta}\rangle$
the corresponding anticommutator spectral function
is simply the single particle spectral function
\bege\label{eq_single_particle_specfun}
S_{\vn{k}\beta\alpha}(E)=\frac{-1}{2\pi}\int_{-\infty}^{\infty} d\,t e^{-\frac{i}{\hbar}E t}
    \langle[
 c_{\vn{k}\beta},c^{\dagger}_{\vn{k}\alpha}(t)
    ]_{+}\rangle,    
\ee
which
we discuss above in the preceding section.
However,
\bege\label{eq_anticommutator_spec_abcd}
S_{l\gamma m\delta j\beta \vn{k}\alpha}(E)=
\int_{-\infty}^{\infty}d\,t
\frac{-e^{-\frac{i}{\hbar}E t}}{2\pi}
\langle[c^{\dagger}_{j\beta} c_{l\gamma} c_{m\delta},c^{\dagger}_{\vn{k}\alpha}(t)]_{+} \rangle
\ee
in Eq.~\eqref{eq_corr4_specfun}
still needs to be determined.

For this we may use again
the method of spectral moments, i.e.,
we compute
\bege
K^{(n)}_{l\gamma m\delta j\beta \vn{k}\alpha}=\frac{1}{\hbar}
\int d\,E [E-\mu]^{n} S_{l\gamma m\delta j\beta \vn{k}\alpha}(E-\mu)
\ee
for $n=0$ and $n=1$ and require them to be equal to
\bege
\tilde{K}^{(0)}_{l\gamma m\delta j\beta \vn{k}\alpha}=\langle[c^{\dagger}_{j\beta} c_{l\gamma} c_{m\delta},c^{\dagger}_{\vn{k}\alpha}]_{+} \rangle,
\ee
and
\bege
\tilde{K}^{(1)}_{l\gamma m\delta j\beta \vn{k}\alpha}=\langle[c^{\dagger}_{j\beta} c_{l\gamma} c_{m\delta},[\mathcal{H},c^{\dagger}_{\vn{k}\alpha}]_{-}]_{+} \rangle,
\ee
respectively.
We need an approximation of $S_{l\gamma m\delta j\beta \vn{k}\alpha}(E)$
in analogy to Eq.~\eqref{eq_spectral_matrix_general}.
The poles of $S_{l\gamma m\delta j\beta \vn{k}\alpha}(E)$
arise from the time-dependence of the operator $c^{\dagger}_{\vn{k}\alpha}(t)$,
i.e., the poles are simply the energies $E_{\vn{k}\gamma p}$ discussed
in the previous section.
For each of these $2 N_{\rm W}$ poles a prefactor needs to be determined.
When $\alpha, \beta, \gamma, \delta$ run from $1,\dots, N_{\rm W}$
the number of prefactors is $2 N_{\rm W}^5$. However, there are only
$2 N_{\rm W}^4$ equations, if we use the first two moments.
This suggests that we need to make use of $\vn{\mathcal{V}}_{\vn{k}\gamma p}$ as well in order
to arrive at an equality between the number of available equations
and the number of unknown coefficients.

Therefore, we approximate $S_{l\gamma m\delta j\beta \vn{k}\alpha}(E)$ by
\bege
\frac{S_{l\gamma m\delta j\beta \vn{k}\alpha}(E-\mu)}{\hbar}=
\sum_{\gamma' p'}
\delta(E-E_{\vn{k}\gamma' p'})
\mathcal{V}_{\vn{k}\alpha \gamma' p'}
a_{l\gamma m\delta j\beta}^{(\vn{k}\gamma'p')}.
\ee
When we fix the orbital indices $\gamma,\delta,\beta$,
the site-indices $l,m,j$, and the
$\vn{k}$-point $\vn{k}$
we obtain $2N_{\rm W}$
equations if we consider the first two moments and
compute these moments for $\alpha=1,N_{\rm W}$.
The number of unknown coefficients $a_{l\gamma m\delta j\beta}^{(\vn{k}\gamma'p')}$
is likewise $2 N_{\rm W}$,
because $(\gamma'p')$ may take $2 N_{\rm W}$ possible values.
Thus, the
requirement $K^{(n)}_{l\gamma m\delta j\beta \vn{k}\alpha}=\tilde{K}^{(n)}_{l\gamma m\delta j\beta \vn{k}\alpha}$
provides $2 N_{\rm W}$ linear equations for $2 N_{\rm W}$
unknowns $a_{l\gamma m\delta j\beta}^{(\vn{k}\gamma'p')}$.

The evaluation of the anticommutators
\bege
\tilde{K}^{(1)}_{l\gamma m\delta j\beta \vn{k}\alpha}=\langle[c^{\dagger}_{j\beta} c_{l\gamma} c_{m\delta},[\mathcal{H},c^{\dagger}_{\vn{k}\alpha}]_{-}]_{+} \rangle
\ee
may require
correlation functions of the
type $\langle c^{\dagger}_{i'\alpha'}  c^{\dagger}_{j'\beta'} c_{l'\gamma'} c_{m'\delta'} \rangle$.
Therefore, the calculation of the correlation
functions $\langle c^{\dagger}_{i\alpha}  c^{\dagger}_{j\beta} c_{l\gamma} c_{m\delta} \rangle$
has to be performed self-consistently, as explained in the previous section.
In Appendix~\ref{sec_appendix_correlators} we provide examples
for $\tilde{K}^{(n)}_{l\gamma m\delta j\beta \vn{k}\alpha}$.

\section{Response functions}
\subsection{Independent electrons}
\label{sec_indep_electrons}
It is instructive to consider first the derivation of
response functions of independent electron systems using the
method of spectral moments, because the standard procedure~\cite{bastin_conductivity,crepieux_bruno_ahe}
of evaluating the Kubo formula 
does not make use of this method.
Considering independent electrons first will allow us to
establish the basic steps needed to evaluate the response
functions based on the method of spectral moments for correlated
electrons in the next section.

To be concrete, we consider the response function  of the
anomalous Hall effect, i.e., the AHE conductivity $\sigma_{xy}$,
which is given by
\bege\label{eq_ahe_conduc_from_green}
\sigma_{xy}=-\frac{e^2}{NV} \lim_{E \to 0}
\sum_{\vn{k}}\frac{{\rm Im}  G^{\rm R}_{\vn{k}v_x v_y}(E) }{E}
\ee
in terms of the retarded velocity-velocity
correlation function $G^{\rm R}_{v_x v_y}(E)$,
which is defined by
\bege
G^{\rm R}_{\vn{k}v_x v_y}(E)=-i\int_{0}^{\infty}d\,t e^{iEt/\hbar}
\langle
    [
v_x(t),v_y(0)
      ]_{-}
\rangle.
\ee
Here, $v_x$ and $v_y$ are the $x$ and $y$ components of the
velocity operator, respectively, $e$ is the elementary
charge, $V$ is the volume of the unit cell, and $N$ is the
number of $\vn{k}$ points.

One standard derivation of the AHE conductivity
uses Wick's theorem in order to evaluate $G^{\rm R}_{\vn{k}v_x v_y}(E)$ 
for independent electrons.
In order to use the method of spectral moments, one instead needs to
employ the spectral
representation~\cite{book_Nolting}
\bege\label{eq_spec_rep}
G^{\rm R}_{\vn{k}v_x v_y}(E)=\int_{-\infty}^{\infty} d\,E'
\frac{Z_{\vn{k}v_x v_y}(E')}{E-E'+i0^{+}}
\ee
to express $G^{\rm R}_{\vn{k}v_x v_y}(E)$ in terms of the
corresponding spectral function
\bege\label{eq_spec_func_indep}
Z_{\vn{k}v_x v_y}(t-t')=\sum_{\gamma}
\frac{f_{\vn{k}\gamma}}{2\pi}
\langle
\psi_{\vn{k}\gamma}|
    [v_x(t),v_y(t')]_{-}
    |
    \psi_{\vn{k}\gamma}
\rangle
\ee
with Fourier transform
\bege
Z_{\vn{k}v_x v_y}(E)=\int_{-\infty}^{\infty}d\,t e^{i E t/\hbar}Z_{\vn{k}v_x v_y}(t),
\ee
where $f_{\vn{k}\gamma}$ is the Fermi factor for
eigenstate $\gamma$ at $k$-point $\vn{k}$.
$Z_{\vn{k}v_x v_y}(t-t')$ is a commutator spectral function
in contrast to the single particle spectral function
Eq.~\eqref{eq_single_particle_specfun}.

For independent electrons,
the first two commutator moments are easy to
evaluate:
\bege
\begin{aligned}
&\tilde{D}^{(0)}_{\vn{k}v_{x}v_{y}}=\sum_{\gamma}f_{\vn{k}\gamma}
\langle
\psi_{\vn{k}\gamma} |
    [v_{x},v_{y}]_{-}
    |\psi_{\vn{k}\gamma}\rangle\\
    &=\sum_{\gamma\gamma'}
    [f_{\vn{k}\gamma}
      -f_{\vn{k}\gamma'}
      ]
\langle
\psi_{\vn{k}\gamma} |
v_{x}
|\psi_{\vn{k}\gamma'}\rangle
\langle
\psi_{\vn{k}\gamma'} |
v_{y}    
|\psi_{\vn{k}\gamma}\rangle,    
\\  
\end{aligned}  
\ee
and
\bege
\begin{aligned}
&\tilde{D}^{(1)}_{\vn{k}v_{x}v_{y}}=\sum_{\gamma}f_{\vn{k}\gamma}
\langle
\psi_{\vn{k}\gamma} |
    [[v_{x},\mathcal{H}]_{-},v_{y}]_{-}
    |\psi_{\vn{k}\gamma}\rangle=\\
    &=\sum_{\gamma\gamma'}
[
    f_{\vn{k}\gamma}
-f_{\vn{k}\gamma'}
    ]
[
E_{\vn{k}\gamma'}-E_{\vn{k}\gamma}
]\times\\
&
\times\langle
\psi_{\vn{k}\gamma} |
v_{x}
|\psi_{\vn{k}\gamma'}\rangle
\langle
\psi_{\vn{k}\gamma'} |
v_{y}    
|\psi_{\vn{k}\gamma}\rangle.\\  
\end{aligned}  
\ee
Clearly, the spectral function
\bege\label{eq_spec_func_guess_ahe}
\begin{aligned}
&Z_{\vn{k}v_{x}v_{y}}(E)=
\hbar\sum_{\gamma\gamma'}
        [f_{\vn{k}\gamma}-f_{\vn{k}\gamma'}]\times\\
&\times\delta(E-(E_{\vn{k}\gamma'}-E_{\vn{k}\gamma}))
\langle
\psi_{\vn{k}\gamma} |
v_{x}
|\psi_{\vn{k}\gamma'}\rangle
\langle
\psi_{\vn{k}\gamma'} |
v_{y}    
|\psi_{\vn{k}\gamma}\rangle   
\end{aligned}    
\ee
reproduces these first two moments.
Using Eq.~\eqref{eq_spec_rep}
we get the corresponding
retarded Green function
\bege\label{eq_green_final}
G^{\rm R}_{\vn{k}v_x v_y}(E)=\hbar\sum_{\gamma\gamma'}
[f_{\vn{k}\gamma}-f_{\vn{k}\gamma'}]
\frac{\langle
\psi_{\vn{k}\gamma} |
v_{x}
|\psi_{\vn{k}\gamma'}\rangle
\langle
\psi_{\vn{k}\gamma'} |
v_{y}    
|\psi_{\vn{k}\gamma}\rangle}
{E-(E_{\vn{k}\gamma'}- E_{\vn{k}\gamma}) +i0^{+}}.
\ee
Plugging Eq.~\eqref{eq_green_final} into
Eq.~\eqref{eq_ahe_conduc_from_green}
we obtain the
literature expression for the
intrinsic AHE
conductivity~\cite{PhysRevLett.92.037204,RevModPhys.82.1539}:
\bege\label{eq_ahe_conduc_lit}
\begin{aligned}
\sigma_{xy}=&
\sum_{\vn{k}\gamma\gamma'}
\frac{e^2\hbar[f_{\vn{k}\gamma}\!-\! f_{\vn{k}\gamma'}]}{ V N}\times\\
&\times\frac{
{\rm Im}\left[
      \langle
\psi_{\vn{k}\gamma} |
v_{x}
|\psi_{\vn{k}\gamma'}\rangle
\langle
\psi_{\vn{k}\gamma'} |
v_{y}    
|\psi_{\vn{k}\gamma}\rangle\right]}{(E_{\vn{k}\gamma'}- E_{\vn{k}\gamma})^2+0^{+}}.
\end{aligned}
\ee

Obviously, the derivation above is not entirely satisfactory,
because we used only two moments, namely $D^{(0)}_{\vn{k}v_{x}v_{y}}$
and $D^{(1)}_{\vn{k}v_{x}v_{y}}$ in order to guess the spectral function
in Eq.~\eqref{eq_spec_func_guess_ahe}, which contains
up to $N_{\rm W} (N_{\rm W}-1)+1$ different poles. In order to
derive the transition rates for $N_{\rm W} (N_{\rm W}-1)+1$ poles
rigorously
instead of simply guessing them requires the same number of equations,
but the number of moments $D^{(n)}_{\vn{k}v_{x}v_{y}}$ that we can easily
compute will generally be much smaller than the number of these poles.
Therefore, we consider instead the moments
\bege\label{eq_moment_D0}
\tilde{D}^{(0)}_{\vn{k}\alpha \beta \gamma \delta}=
\langle
    [c^{\dagger}_{\vn{k}\alpha}c_{\vn{k}\beta},c^{\dagger}_{\vn{k}\gamma}c_{\vn{k}\delta}]_{-}
\rangle,
\ee
and
\bege\label{eq_moment_D1}
\tilde{D}^{(1)}_{\vn{k}\alpha \beta \gamma \delta}=
\langle
    [[c^{\dagger}_{\vn{k}\alpha}c_{\vn{k}\beta},\mathcal{H}]_{-},c^{\dagger}_{\vn{k}\gamma}c_{\vn{k}\delta}]_{-}
\rangle.
\ee
The moments $\tilde{D}^{(n)}_{\vn{k}v_{x}v_{y}}$ considered so far
are simply contractions of these new moments
with the velocity operators:
\bege
\sum_{\alpha\beta\gamma\delta}\tilde{D}^{(n)}_{\vn{k}\alpha \beta \gamma \delta}v_{x\alpha\beta}v_{y\gamma\delta}=
\tilde{D}^{(n)}_{\vn{k}v_{x}v_{y}}.
\ee
The indices $\alpha$, $\beta$, $\gamma$, and $\delta$
may be the band indices of the eigenstates $|\psi_{\vn{k}\alpha}\rangle$,
but they may also be the indices used to label the
basis functions $|\phi_{\vn{k}\alpha}\rangle$. We do not introduce different
notations for labelling eigenstates on the one hand and basis
states on the other hand. The basis state representation and the
eigenstate representation are connected by the unitary transformation
Eq.~\eqref{eq_unitary_trafo}.

In the eigenstate representation we obtain
particularly simple expressions for these moments:
\bege\label{eq_d0_eigenstate}
\tilde{D}^{(0)}_{\vn{k}\alpha \beta \gamma \delta}=
[f_{\vn{k}\alpha}-f_{\vn{k}\beta}]\delta_{\alpha\delta}\delta_{\gamma\beta}
\ee
and
\bege\label{eq_d1_eigenstate}
\tilde{D}^{(1)}_{\vn{k}\alpha \beta \gamma \delta}=
[f_{\vn{k}\alpha}-f_{\vn{k}\beta}][E_{\vn{k}\beta}-E_{\vn{k}\alpha}]\delta_{\alpha\delta}\delta_{\gamma\beta}.
\ee
Using these two spectral moments we may
easily obtain the corresponding spectral function
\bege\label{eq_specfunc_abcd_timedep}
Z_{\vn{k}\alpha\beta\gamma\delta}(t-t')=
\frac{\langle
[c^{\dagger}_{\vn{k}\alpha}(t)c_{\vn{k}\beta}(t),c^{\dagger}_{\vn{k}\gamma}(t')c_{\vn{k}\delta}(t')]_{-}
\rangle}{2\pi}
\ee
as
\bege\label{eq_specmom_indep_rig}
\frac{Z_{\vn{k}\alpha\beta\gamma\delta}(E)}{\hbar}
=[f_{\vn{k}\alpha}-f_{\vn{k}\beta}]\delta_{\alpha\delta}\delta_{\gamma\beta}
\delta(E-(E_{\vn{k}\beta}-E_{\vn{k}\alpha})).
\ee
Employing
\bege\label{eq_svv_from_scccc}
Z_{\vn{k}v_{x}v_{y}}(E)=
\sum_{\alpha\beta\gamma\delta}
Z_{\vn{k}\alpha \beta \gamma \delta}(E)
v_{x\alpha\beta}v_{y\gamma\delta}
\ee
we obtain Eq.~\eqref{eq_spec_func_guess_ahe}
from Eq.~\eqref{eq_specmom_indep_rig}, which completes the rigorous
derivation of the AHE conductivity of independent electrons based
on the method of spectral moments.
Certainly, Eq.~\eqref{eq_spec_func_indep} may also be evaluated easily
directly for independent electrons, without using the method of
spectral moments. However, the purpose of this section is to
establish the necessary guiding principles to find the commutator spectral function
of correlated electrons in the following section.

Using the unitary transformation Eq.~\eqref{eq_unitary_trafo}
we may transform
the spectral moments and the commutator
spectral function given in the eigenstate representations
in Eq.~\eqref{eq_d0_eigenstate}, Eq.~\eqref{eq_d1_eigenstate},
and Eq.~\eqref{eq_specmom_indep_rig}
into the representation of basis functions:
\bege
\begin{aligned}
\tilde{D}^{(n)}_{\vn{k}\alpha'\beta'\gamma'\delta'}&=
\sum_{\alpha\beta}
    [f_{\vn{k}\alpha}-f_{\vn{k}\beta}]
    [E_{\vn{k}\beta}-E_{\vn{k}\alpha}]^{n}\times \\
    &\times
    \mathcal{U}_{\vn{k}\alpha'\alpha}
    \mathcal{U}^{*}_{\vn{k}\beta'\beta}
    \mathcal{U}_{\vn{k}\gamma'\beta}
    \mathcal{U}^{*}_{\vn{k}\delta'\alpha},
\end{aligned}    
\ee
and
\bege\label{eq_commu_spec_nonint}
\begin{aligned}
\frac{Z_{\vn{k}\alpha'\beta'\gamma'\delta'}(E)}{\hbar}&=
\sum_{\alpha\beta}
    [f_{\vn{k}\alpha}-f_{\vn{k}\beta}]
    \delta(E-(E_{\vn{k}\beta}-E_{\vn{k}\alpha}))\times \\
    &\times
    \mathcal{U}_{\vn{k}\alpha'\alpha}
    \mathcal{U}^{*}_{\vn{k}\beta'\beta}
    \mathcal{U}_{\vn{k}\gamma'\beta}
    \mathcal{U}^{*}_{\vn{k}\delta'\alpha}.
\end{aligned}    
\ee

\subsection{Correlated electrons}
\label{sec_resp_corr}
When we treat the correlated electron system
with the method of Sec.~\ref{sec_ground_state}
there are $N_{\rm W}$ basis functions but up to $2 N_{\rm W}$
energy bands due to the splitting into the lower and the
upper Hubbard band. Consequently, the labels
$\alpha$, $\beta$, $\gamma$, $\delta$ in
Eq.~\eqref{eq_moment_D0}, Eq.~\eqref{eq_moment_D1},
and Eq.~\eqref{eq_specfunc_abcd_timedep} refer now
only to the basis functions and not to the
energy bands, i.e., $\alpha,\beta,\gamma,\delta=1,\dots N_{\rm W}$.

In general, the commutator spectral function
exhibits the following properties:
It is hermitean, i.e.,
\bege\label{eq_hermit_commut_spec}
[Z_{\alpha\beta\gamma\delta}(E)]^{*}=Z_{\delta\gamma\beta\alpha}(E),
\ee
and it 
fulfills 
\bege\label{eq_commut_spec_negene}
Z_{\vn{k}\alpha\beta\gamma\delta}(E)=-Z_{\vn{k}\gamma\delta\alpha\beta}(-E).
\ee
One may additionally combine Eq.~\eqref{eq_hermit_commut_spec}
and Eq.~\eqref{eq_commut_spec_negene}
to obtain
\bege\label{eq_require_minusE}
Z_{\vn{k}\alpha\beta\gamma\delta}(E)=-[Z_{\vn{k}\beta\alpha\delta\gamma}(-E)]^{*}.
\ee

The differences between the single particle spectral functions
of correlated electrons on the one hand (Eq.~\eqref{eq_spectral_matrix_general}) 
and independent electrons on the other hand (Eq.~\eqref{eq_spectral_matrix_nonint})
are the additional spectral weights $a_{\vn{k}\gamma p}$
and the replacement of the unitary matrix $\vn{\mathcal{U}}_{\vn{k}}$
by the matrix $\vn{\mathcal{V}}_{\vn{k}}$ of the state vectors.
It is therefore plausible to guess the commutator spectral function
of correlated electrons
by adding spectral weights to Eq.~\eqref{eq_commu_spec_nonint}
and by replacing $\vn{\mathcal{U}}_{\vn{k}}$
by $\vn{\mathcal{V}}_{\vn{k}}$, which yields
\bege\label{eq_commu_spec_int_guess}
\begin{aligned}
&Z_{\vn{k}\alpha\beta\gamma\delta}(E)=\hbar
\sum_{p,p'=1}^{2}
\sum_{\mu,\mu'=1}^{N_{\rm W}}
    \delta(E-(E_{\vn{k}\mu' p'}-E_{\vn{k}\mu p}))\times \\
    &\times
    a_{\vn{k}\mu' p'}a_{\vn{k}\mu p}
[f_{\vn{k}\mu p}-f_{\vn{k}\mu' p'}]
    \mathcal{V}_{\vn{k}\alpha\mu p}\mathcal{V}^{*}_{\vn{k}\delta\mu p}
    \mathcal{V}^{*}_{\vn{k}\beta\mu' p'}\mathcal{V}_{\vn{k}\gamma\mu' p'}.
\end{aligned}    
\ee

In order to test the quality of this guess 
Eq.~\eqref{eq_commu_spec_int_guess}
we may compute the first two moments from it, i.e., evaluate the
energy integrals similar to Eq.~\eqref{eq_specmoms_eneint}
(see Eq.~\eqref{eq_moment_d_eneint} below), and subsequently 
we may compare these moments to those obtained from evaluating the
alternative commutator expressions Eq.~\eqref{eq_moment_D0}
and Eq.~\eqref{eq_moment_D1}.
The zeroth moment obtained by integrating Eq.~\eqref{eq_commu_spec_int_guess}
over the energy
is simply
\bege
\begin{aligned}
&D^{(0)}_{\vn{k}\alpha\beta\gamma\delta}=
\sum_{p,p'=1}^{2}
\sum_{\mu,\mu'=1}^{N_{\rm W}}
    a_{\vn{k}\mu' p'}a_{\vn{k}\mu p}
[f_{\vn{k}\mu p}-f_{\vn{k}\mu' p'}]\times\\
    &\times\mathcal{V}_{\vn{k}\alpha\mu p}\mathcal{V}^{*}_{\vn{k}\delta\mu p}
\mathcal{V}^{*}_{\vn{k}\beta\mu' p'}\mathcal{V}_{\vn{k}\gamma\mu' p'}=\\
&=\sum_{p=1}^{2}\sum_{\mu=1}^{N_{\rm W}}a_{\vn{k}\mu p}f_{\vn{k}\mu p}
[
  \mathcal{V}_{\vn{k}\alpha\mu p}\mathcal{V}^{*}_{\vn{k}\delta\mu p}\delta_{\gamma\beta}
-\mathcal{V}^{*}_{\vn{k}\beta\mu p}\mathcal{V}_{\vn{k}\gamma\mu p}\delta_{\alpha\delta}
].  
\end{aligned} 
\ee
The commutator expression in Eq.~\eqref{eq_moment_D0}
evaluates to
\bege\label{eq_d0_commut1}
\begin{aligned}
\tilde{D}^{(0)}_{\vn{k}\alpha\beta\gamma\delta}=\langle c^{\dagger}_{\vn{k}\alpha} c_{\vn{k}\delta} \rangle\delta_{\beta\gamma}-\langle c^{\dagger}_{\vn{k}\gamma} c_{\vn{k}\beta} \rangle\delta_{\alpha\delta},
\end{aligned}  
\ee
where we made use of the identity
\bege
[\hat{A},\hat{B}\hat{C}]_{-}=[\hat{A},\hat{B}]_{+}\hat{C}-\hat{B}[\hat{A},\hat{C}]_{+}
\ee
(where $\hat{A}$, $\hat{B}$ and $\hat{C}$ are operators)
to convert the commutator into anticommutators, which
are simply given by
\bege
\begin{aligned}
  &[c^{\dagger}_{\vn{k}\alpha}, c_{\vn{k}\beta}]_{+}=\delta_{\alpha\beta},\\
  &[c_{\vn{k}\alpha}, c_{\vn{k}\beta}]_{+}=0,\\
  &[c^{\dagger}_{\vn{k}\alpha}, c^{\dagger}_{\vn{k}\beta}]_{+}=0,  
\end{aligned}
\ee
for fermionic creation and annihilation operators.
Inserting
\bege
\langle c^{\dagger}_{\vn{k}\alpha} c_{\vn{k}\beta} \rangle=
\sum_{p=1}^{2}\sum_{\mu=1}^{N_{\rm W}}a_{\vn{k}\mu p}f_{\vn{k}\mu p}
  \mathcal{V}^{*}_{\vn{k}\alpha\mu p}\mathcal{V}_{\vn{k}\beta\mu p},
\ee
which may be derived easily using the spectral theorem
Eq.~\eqref{eq_spec_theo_single_part},
into Eq.~\eqref{eq_d0_commut1},
we obtain $\tilde{D}^{(0)}_{\vn{k}\alpha\beta\gamma\delta}=D^{(0)}_{\vn{k}\alpha\beta\gamma\delta}$,
i.e., our guess Eq.~\eqref{eq_commu_spec_int_guess}
describes the zeroth moment consistently.

The first
moment $D^{(1)}_{\vn{k}\alpha\beta\gamma\delta}=\frac{1}{\hbar}\int d E \mathcal{Z}_{\vn{k}\alpha\beta\gamma\delta}(E) E$
is given by
\bege\label{eq_first_mom_corr_eneint}
\begin{aligned}
&D^{(1)}_{\vn{k}\alpha\beta\gamma\delta}=
\sum_{p,p'=1}^{2}
\sum_{\mu,\mu'=1}^{N_{\rm W}}
(E_{\vn{k}\mu' p'}-E_{\vn{k}\mu p})
a_{\vn{k}\mu' p'}\times \\
&\times a_{\vn{k}\mu p}
[f_{\vn{k}\mu p}-f_{\vn{k}\mu' p'}]
    \mathcal{V}_{\vn{k}\alpha\mu p}\mathcal{V}^{*}_{\vn{k}\delta\mu p}
    \mathcal{V}^{*}_{\vn{k}\beta\mu' p'}\mathcal{V}_{\vn{k}\gamma\mu' p'}.
\end{aligned}    
\ee
Formally, the alternative commutator expressions Eq.~\eqref{eq_moment_D1}
differ considerably from Eq.~\eqref{eq_first_mom_corr_eneint}, while
they yield similar numerical results for the model considered in
Sec.~\ref{sec_rashba_hubbard}.
In Appendix~\ref{app_ahe_moments} we give some examples
of these first moments obtained by evaluating the
commutator expressions Eq.~\eqref{eq_moment_D1}.

While the numerical differences
between $\tilde{D}^{(1)}_{\vn{k}\alpha\beta\gamma\delta}$
and $D^{(1)}_{\vn{k}\alpha\beta\gamma\delta}$
are sufficiently small to consider Eq.~\eqref{eq_commu_spec_int_guess}
as a useful approximation, 
one may obtain almost perfect agreement
between $\tilde{D}^{(1)}_{\vn{k}\alpha\beta\gamma\delta}$
and $D^{(1)}_{\vn{k}\alpha\beta\gamma\delta}$ by
simple modifications of Eq.~\eqref{eq_commu_spec_int_guess}.
One possible improvement of Eq.~\eqref{eq_commu_spec_int_guess}
is the replacement of the two known spectral weight factors
by a single more general unknown coefficient matrix:
\bege\label{eq_trans_weight}
a_{\vn{k}\mu' p'}a_{\vn{k}\mu p}\rightarrow a_{\vn{k}\mu' p'\mu p }.
\ee
This unknown coefficient matrix may be determined by
minimizing the deviation
\bege\label{eq_devi_lambda}
\lambda=
\sum_{\alpha\beta\gamma\delta}\sum_{n=1}^{2}
|\tilde{D}^{(n)}_{\vn{k}\alpha\beta\gamma\delta}-D^{(n)}_{\vn{k}\alpha\beta\gamma\delta}|^{2}.
\ee
The generalization Eq.~\eqref{eq_trans_weight}
removes the problem of Eq.~\eqref{eq_commu_spec_int_guess}
that the state $\mu p$ always contributes with the
weight $a_{\vn{k}\mu p}$ irrespective
of the partner state $\mu' p'$ of the transition.
In order to satisfy Eq.~\eqref{eq_hermit_commut_spec},
Eq.~\eqref{eq_commut_spec_negene}, and
Eq.~\eqref{eq_require_minusE},
these generalized transition weights need to be real-valued
and they have to satisfy
\bege\label{eq_transmod_condi}
a_{\vn{k}\mu' p'\mu p }=a_{\vn{k}\mu p \mu' p'}.
\ee

In Sec.~\ref{sec_ground_state}
and Sec.~\ref{sec_corr_fun}
we solved systems of coupled equations in order to
determine anticommutator spectral functions.
Therefore, the question arises if we may also obtain an expression for the
commutator spectral function by solving a system of coupled
equations instead of starting with the guess Eq.~\eqref{eq_commu_spec_int_guess}
and successively improving it through the modification Eq.~\eqref{eq_trans_weight}.
The expression Eq.~\eqref{eq_specmom_indep_rig}
of the commutator spectral function
of independent electrons suggests the following
form for the commutator spectral function
of correlated electrons:
\bege\label{eq_commu_spec_corr_general}
\begin{aligned}
&Z_{\vn{k}\alpha\beta\gamma\delta}(E)=
\sum_{p,p'=1}^{2}
\sum_{\mu,\mu'=1}^{N_{\rm W}}
    \delta(E-(E_{\vn{k}\mu' p'}-E_{\vn{k}\mu p}))b^{\mu p \mu' p'}_{\alpha\beta\gamma\delta}.
\end{aligned}
\ee
For fixed $n$ the moment
\bege\label{eq_moment_d_eneint}
D^{(n)}_{\alpha\beta\gamma\delta}=\frac{1}{\hbar}\int_{-\infty}^{\infty}Z_{\vn{k}\alpha\beta\gamma\delta}(E) E^{n}d\,E
\ee
has $N_{\rm W}^4$ components. Consequently, considering the first two moments
provides $2 N_{\rm W}^4 $ equations to determine the unknown
parameters $b^{\mu p \mu' p'}_{\alpha\beta\gamma\delta}$.
However, the number of components $b^{\mu p \mu' p'}_{\alpha\beta\gamma\delta}$
is $4N_{\rm W}^6$. 
Without imposing additional constraints on the form of $b^{\mu p \mu' p'}_{\alpha\beta\gamma\delta}$
we therefore do not have a sufficient number of equations to
determine $b^{\mu p \mu' p'}_{\alpha\beta\gamma\delta}$.

In Sec.~\ref{sec_corr_fun}
above we describe a similar difficulty for the anticommutator spectral function,
which we solve by using additionally the state vectors $\vn{\mathcal{V}}_{\vn{k}\gamma p}$
in the approximation
of the spectral function. Therefore, it is plausible to make use of
both $E_{\vn{k}\mu p}$ and $\mathcal{V}_{\vn{k}\gamma p}$ to formulate a suitable approximation for the
commutator spectral function. However, this is what we describe above already.
The main difference between our solution above
and the previous solution strategies in Sec.~\ref{sec_ground_state} and Sec.~\ref{sec_corr_fun}
is that we minimize Eq.~\eqref{eq_devi_lambda} instead of solving systems of coupled
equations, because there is no obvious modification of
Eq.~\eqref{eq_commu_spec_corr_general} which requires as many unknown parameters
as the number of equations provided by the first two moments.
The number of unknown coefficients $a_{\vn{k}\mu' p'\mu p }$
is
$N_{\rm W}(2N_{\rm W}-1)$
in total. Since the moments are hermitean, the first two moments
may be expressed in terms of $2 N_{\rm W}^4 $ real numbers. This number is sufficient
to obtain  $a_{\vn{k}\mu' p'\mu p }$
unambiguously
by minimizing the deviation $\lambda$ in Eq.~\eqref{eq_devi_lambda}.

Using Eq.~\eqref{eq_svv_from_scccc},
Eq.~\eqref{eq_ahe_conduc_from_green},
and Eq.~\eqref{eq_spec_rep}
we obtain the following expression for the AHE conductivity
from Eq.~\eqref{eq_commu_spec_int_guess} and
Eq.~\eqref{eq_trans_weight}:
\bege\label{eq_ahe_conduct}
\begin{aligned}
&\sigma_{xy}=\frac{e^2\hbar}{ V N}
\sum_{p,p'=1}^{2}
\sum_{\mu,\mu'=1}^{N_{\rm W}}
a_{\vn{k}\mu' p'\mu p}
[f_{\vn{k}\mu p}-f_{\vn{k}\mu' p'}]\times\\
&\times
\sum_{\alpha\beta\gamma\delta}
\frac{{\rm Im}\left [v_{x \alpha\beta}v_{y \gamma \delta}
\mathcal{V}_{\vn{k}\alpha\mu p}\mathcal{V}^{*}_{\vn{k}\delta\mu p}
\mathcal{V}^{*}_{\vn{k}\beta\mu' p'}\mathcal{V}_{\vn{k}\gamma\mu' p'}\right]}
{(E_{\vn{k}\mu' p'}-E_{\vn{k}\mu p})^2+0^{+}}.
\end{aligned}    
\ee

\section{Application to the Hubbard-Rashba model}
\label{sec_rashba_hubbard}
In Sec.~\ref{sec_ground_state} we explained that $4N_{\rm W}^2$
unknown coefficients need to be determined in order to obtain
the spectral function.
While the one-band Hubbard model has spin-up and spin-down bands,
one may obtain these bands separately when there is no spin-orbit interaction (SOI).
Therefore, effectively $N_{\rm W}=1$ for the one-band Hubbard model
without SOI, i.e.,  $4N_{\rm W}^2$ evaluates to
4. This is indeed the number of
parameters in the two-pole approximation Eq.~\eqref{eq_twopole_twoband}.
Without SOI, one computes the spin-up and
spin-down bands separately and 4 parameters are needed for each of them.

However, when we add SOI
to the one-band Hubbard model the spin-up and
spin-down bands are coupled. Consequently, 
we need to use $N_{\rm W}=2$, and $4N_{\rm W}^2=16$ is the
number of unknown coefficients. Therefore,
we may demonstrate the spectral moment approach 
with noncommuting spectral moment matrices
developed in this paper for the single-band Hubbard model with 
additional Rashba-type SOI, because the standard two-pole approximation
cannot be used in this case.

\begin{figure}
\includegraphics[angle=0,width=\linewidth]{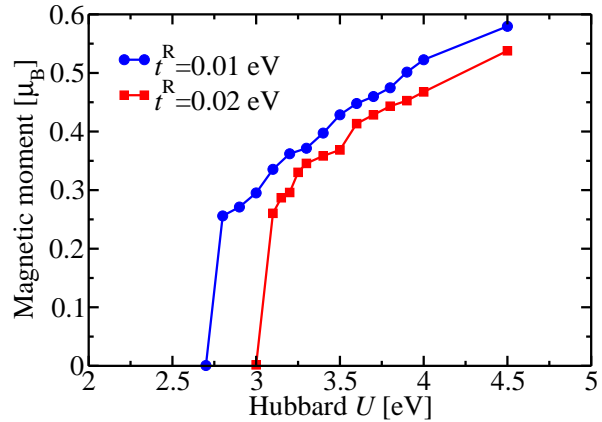}
\caption{\label{fig_magmom_vs_u}
  Dependence of the magnetic moment on the Hubbard parameter $U$
  at temperature $T=23$~K. Two values of SOI strength are considered:
  $t^{\rm R}=0.01$~eV (blue circles)
  and $t^{\rm R}=0.02$~eV (red squares). The magnetic moment
  is given in units of the Bohr magneton ($\mu_{\rm B}$).
}
\end{figure}

We consider the Hubbard-Rashba model~\cite{rashbahub_PhysRevB.88.045102}
with the Hamiltonian
\bege\label{eq_rashba_hubbard}
\begin{aligned}
  H&=\sum_{<j,l>,s}t^{\phandag}_{jl}c^{\dagger}_{j s} c^{\phandag}_{l s}
  +U\sum_{j}c^{\dagger}_{j \uparrow} c^{\phandag}_{j \uparrow}
  c^{\dagger}_{j \downarrow} c^{\phandag}_{j \downarrow}
\\
&+
i t^{\rm R}
\sum_{<j,l>,s,s'}
\hat{\vn{e}}_{z}\cdot (\vn{\sigma}_{ss'} \times \vn{d}^{\phandag}_{jl})
c^{\dagger}_{j s} c^{\phandag}_{l s'},\\
\end{aligned}
\ee
where $t^{\phandag}_{jl}$ is the hopping amplitude, $U$ is the
strength of the Hubbard interaction, 
$t^{\rm R}$ quantifies the Rashba-type SOI,
and $\vn{d}_{jl}=(\vn{R}_{j}-\vn{R}_{l})/a$ is the distance vector
between sites $j$ and $l$, where $a$ is the
lattice constant. The notation $<j,l>$ means that
$j$ and $l$ are nearest neighbors. We consider a two-dimensional
square lattice with lattice translation
invariance and $\hat{\vn{e}}_{z}$
is a unit vector perpendicular to the lattice.

For this model we give explicit expressions for the moments
in Appendix~\ref{sec_appendix_hubbard}. 
The moments needed to evaluate the correlation functions
of Sec.~\ref{sec_corr_fun}
are discussed in Appendix~\ref{sec_appendix_correlators},
while explicit expressions for the moments required for
the calculation of the AHE are given in Appendix~\ref{app_ahe_moments}.

We set the nearest neighbor
hopping
to $t^{\phandag}_{jl}=-0.2{\rm eV}\delta_{1,|d_{xjl}|+|d_{yjl}|}$,
the site occupation
\bege
n=n_{\uparrow}+n_{\downarrow}=\langle c^{\dagger}_{i\uparrow } c_{i\uparrow }\rangle
+\langle c^{\dagger}_{i\downarrow } c_{i\downarrow }\rangle
\ee
to $n=0.675$ and we vary the Hubbard U parameter between
2~eV and 5~eV.
We perform calculations for two SOI strengths, $t^{\rm R}=0.01$~eV,
and $t^{\rm R}=0.02$~eV. We use an $80\times 80$ $k$-mesh in the calculations.

In Fig.~\ref{fig_magmom_vs_u}
we show the magnetic moment as a function of $U$
at the temperature $T=23$~K.
Around $U=2.7$~eV the system becomes ferromagnetic when $t^{\rm R}=0.01$~eV.
When SOI is larger, i.e.,  $t^{\rm R}=0.02$~eV,
the onset of ferromagnetism occurs at a higher $U$ of around 3~eV.
Therefore, SOI suppresses the onset of ferromagnetism in this system.

\begin{figure}
\includegraphics[angle=-90,width=\linewidth]{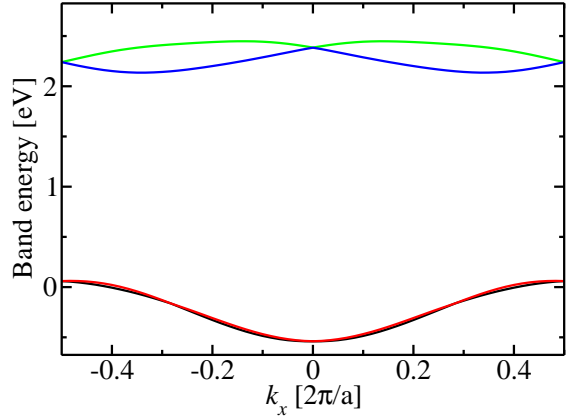}
\caption{\label{fig_U2_nonmag}
  Bandstructure for $U=2$~eV at $T$=23~K
  when $t^{\rm R}=0.02$~eV.
  For this value of $U$ the
  system is non-magnetic. The Rashba splitting is
  larger for the upper Hubbard bands than for the lower ones.
}
\end{figure}

\begin{figure}
\includegraphics[width=\linewidth]{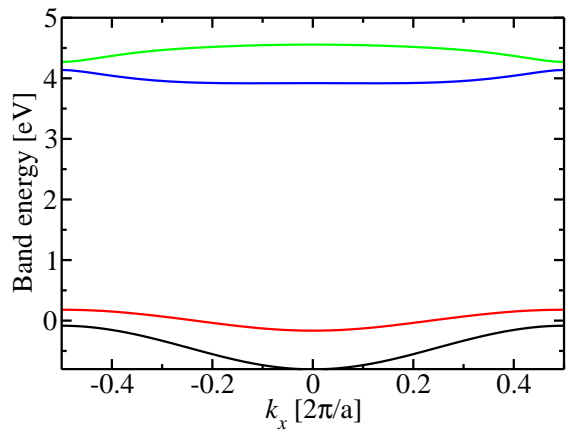}
\caption{\label{fig_U4_ferro}
  Bandstructure for $U=4$~eV at $T$=23~K
  when $t^{\rm R}=0.02$~eV. For this value of $U$ the
  system is ferromagnetic.
}
\end{figure}

In Fig.~\ref{fig_U2_nonmag} we show the bandstructure
in the nonmagnetic phase at $U=2$~eV.
The lower Hubbard bands describe electrons that hop between
empty sites, while the upper Hubbard bands describe electrons
hopping between sites that are already occupied by an electron.
Consequently, the lower and upper Hubbard bands are separated
in energy by roughly $U$.
Surprisingly,
the Rashba splitting is larger for the upper Hubbard bands
than for the lower ones.
The bandstructure in the ferromagnetic phase at $U=4$~eV
is shown in Fig.~\ref{fig_U4_ferro}. Both the upper and the
lower Hubbard bands are exchange-split, while the effect
of Rashba SOI on the bandstructure is not visible directly.

\begin{figure}
\includegraphics[width=\linewidth]{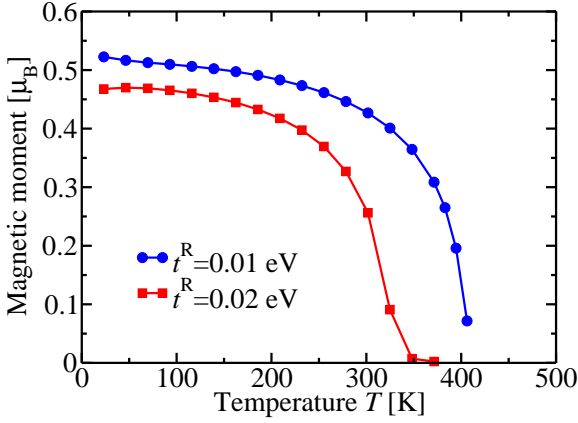}
\caption{\label{fig_U4_magneti_tempdep}
  Temperature-dependence of the magnetic moment at $U=4$~eV for
  SOI-strength $t^{\rm R}=0.01$~eV (circles)
  and $t^{\rm R}=0.02$~eV (squares). Stronger SOI suppresses the
  Curie temperature.
}
\end{figure}

In Fig.~\ref{fig_U4_magneti_tempdep} we show the temperature-dependence
of the magnetic moment at $U=4$~eV. The magnetic moment decreases
monotonously with temperature.
The Curie temperature is smaller when SOI is larger.
This is consistent with the finding in Fig.~\ref{fig_magmom_vs_u}
that SOI suppresses ferromagnetism in this system.

\begin{figure}
\includegraphics[width=\linewidth]{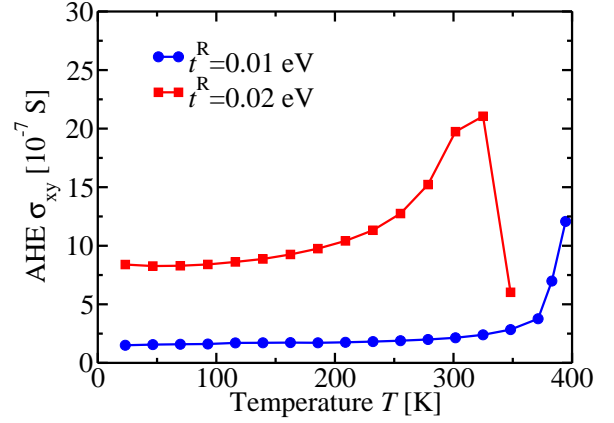}
\caption{\label{fig_U4_AHE_tempdep}
  Temperature-dependence of the intrinsic AHE conductivity at $U=4$~eV. 
}
\end{figure}

In Fig.~\ref{fig_U4_AHE_tempdep} we show the temperature-dependence
of the intrinsic AHE conductivity. Since AHE requires SOI,
higher SOI leads to larger AHE.
Interestingly, $\sigma_{xy}$ increases with
increasing $T$.
Phenomenological theory usually assumes that $\sigma_{xy}$ is proportional
to the magnetization. However, from zero temperature
up to the Curie temperature
the magnetic moment
decreases, while the AHE conductivity increases
according to Fig.~\ref{fig_U4_AHE_tempdep}.
This increase of the AHE conductivity with decreasing magnetic moment
is therefore surprising at first.

An explanation may be found by
recalling in which cases the AHE is predicted to be proportional
to the magnetic moment. When the intrinsic AHE is generated predominantly
by the term $\propto L_{z}S_{z}$ in the SOI ($L_{z}$ and $S_{z}$ are the orbital and spin
angular momentum operators, respectively) and when the magnetic moment is sufficiently small
we may argue that the spin-resolved AHE conductivity  has opposite sign for
spin-up and spin-down electrons, i.e., ${\rm sign}(\sigma^{\uparrow}_{xy})=-{\rm sign}(\sigma^{\downarrow}_{xy})$,
because $L_{z}S_{z}=L_{z}$ for spin-up electrons and $L_{z}S_{z}=-L_{z}$ for spin-down electrons.
As a consequence, a pure spin-current is generated from the intrinsic spin-Hall
effect when the magnetization is
zero: $\sigma^{\rm SHE}_{xy}=\sigma^{\uparrow}_{xy}-\sigma^{\downarrow}_{xy}=2\sigma^{\uparrow}_{xy}$.
For small magnetic moment
the spin-resolved AHE conductivities do not
satisfy $\sigma^{\uparrow}_{xy}=-\sigma^{\downarrow}_{xy}$
any more~\cite{PhysRevB.91.115316}
and it is plausible to assume that both depend linearly on the
magnetization $M$ in the leading orders,
\bege\label{eq_sigma_linear_in_m}
\sigma^{s}_{xy}(M)=s\sigma^{(0)}_{xy}+\left.\frac{\partial \sigma^{(0)}_{xy}}{ \partial M}\right|_{M=0}M,
\ee
where $s=1$ for spin-up and $s=-1$ for spin-down.
This yields
\bege
\begin{aligned}
  &\sigma^{\rm SHE}_{xy}(M)=\sigma^{\uparrow}_{xy}(M)-\sigma^{\downarrow}_{xy}(M)=2\sigma^{(0)}_{xy},\\
  &\sigma^{\rm AHE}_{xy}(M)=\sigma^{\uparrow}_{xy}(M)+\sigma^{\downarrow}_{xy}(M)=2\left.\frac{\partial \sigma^{(0)}_{xy}}{ \partial M}\right|_{M=0}M.\\
\end{aligned}
\ee
Thus, one may expect $\sigma^{\rm AHE}_{xy}(M)\propto M$ when
only the term $L_{z}S_{z}$ from SOI is relevant.
However, this is often not the case, i.e., spin-flip
transitions may be important for the
AHE~\cite{PhysRevLett.106.117202}.

In fact the Rashba SOI does not contain any term that preserves the spin $S_z$,
i.e., in the language of Ref.~\cite{PhysRevLett.106.117202} it generates
only spin-flip transitions.
However, when all virtual transitions that give rise to the
AHE are of spin-flip type, there is no
reason for
Eq.~\eqref{eq_sigma_linear_in_m}
to be valid.
The spin-flip transitions occur between up- and down-states,
and their energy difference decreases when the magnetic moment
decreases. Therefore, the spin-flip transitions become
more important in
Eq.~\eqref{eq_ahe_conduct}
when the energy denominator decreases due to the decrease of the
magnetization.

\begin{figure}
\includegraphics[width=\linewidth]{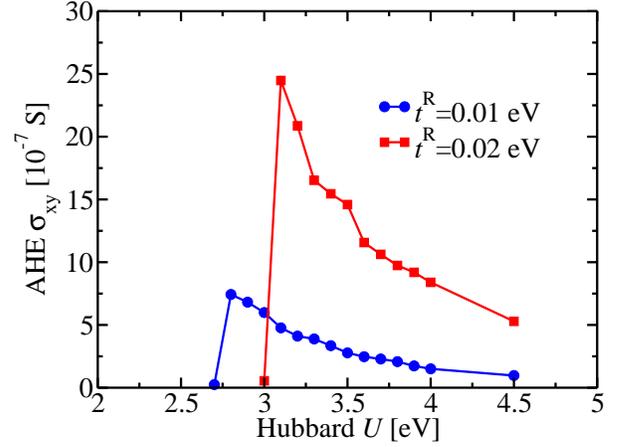}
\caption{\label{fig_AHE_udep}
  Dependence of the intrinsic AHE conductivity on
  Hubbard U at $T=23$~K. 
}
\end{figure}

To verify this hypothesis we investigate the dependence of
the AHE conductivity on the Hubbard parameter U at a fixed
small temperature of $T=23$~K.
Thereby we modify the magnetic moment without changing the temperature.
The result is shown in Fig.~\ref{fig_AHE_udep}.
Indeed $\sigma_{xy}$ increases when lowering $U$ from 4.5~eV down to 2.8~eV.
Therefore, the dominant mechanism for the strong increase of $\sigma_{xy}$
with increasing temperature in Fig.~\ref{fig_U4_AHE_tempdep}
is indeed the decrease of the magnetic moment.

\section{Outlook}
\label{sec_outlook}
In order to judge if the selfconsistent spectral moment
method may replace LDA+$U$ while keeping the computational
effort low, we briefly discuss the differences to LDA+$U$.
The orbital-resolved occupation matrix, which needs to
be computed self-consistently in LDA+$U$ calculations
for all atomic orbitals for which correlations are included,
corresponds to the correlator $\langle c^{\dagger}_{i\alpha} c_{j\beta}\rangle$
in the self-consistent moment method.
In LDA+$U$ only the on-site elements are considered,
i.e., only $\langle c^{\dagger}_{0\alpha} c_{0\beta}\rangle$.
However, including the nearest neighbors will often
not pose problems. Clearly, the matrix structure of
$\langle c^{\dagger}_{i\alpha} c_{j\beta}\rangle$ resembles the one
of the hopping matrix elements, i.e., the matrix elements of
the Hamiltonian between Wannier functions~\cite{wannier90communitycode}.
Therefore, for
all systems where 
the calculation of the hopping matrix does not pose
problems, e.g.\ computer memory issues, we do not anticipate
issues concerning the correlator $\langle c^{\dagger}_{i\alpha} c_{j\beta}\rangle$.

The higher-order correlators of the type
$\langle c^{\dagger}_{i\alpha} c^{\dagger}_{j\beta} c_{l\gamma} c_{m\delta} \rangle$
are often only needed for special cases such as $i=j$ and $l=m$.
Therefore, while the most general
form $\langle c^{\dagger}_{i\alpha} c^{\dagger}_{j\beta} c_{l\gamma} c_{m\delta} \rangle$
has 4 independent site indices, $i,j,l,m$, it will often be sufficient to compute 
it only with one site index. Moreover, when we consider only on-site Coulomb interactions,
similar to standard LDA+$U$, the range of the orbital indices $\alpha,\beta,\gamma,\delta$ 
can be restricted significantly. Thus, while the number of basis functions
in LDA calculations based on FLAPW increases proportionally to the system volume, leading to a quadratic
increase of the memory requirement of the Hamiltonian with system volume, we expect that the computer
memory needed to hold $\langle c^{\dagger}_{i\alpha} c^{\dagger}_{j\beta} c_{l\gamma} c_{m\delta} \rangle$
scales only proportionally to the number of correlated atoms in the system, i.e., often
proportionally to the volume and not to the square of the volume.
This is similar to the scaling of the orbital-resolved occupation matrix with system size
in LDA+$U$, which is proportional
to the number of correlated atoms.

Thus, while in LDA+$U$ one needs to compute the orbital-resolved occupation matrix for all correlated atoms,
in the selfconsistent spectral moment method one has to compute several
correlators of the type $\langle c^{\dagger}_{i\alpha} c_{j\beta}\rangle$
and $\langle c^{\dagger}_{i\alpha} c^{\dagger}_{j\beta} c_{l\gamma} c_{m\delta} \rangle$.
Since their structure and memory requirement is similar to the one of matrix elements computed for Wannier
interpolation~\cite{wannier90communitycode,PhysRevB.74.195118,ibcsoit},
we expect that this is feasible for a large range of realistic materials without
a dramatic increase of the computer time or memory requirements, because experience with
Wannier functions suggests that the calculation of such matrix elements is much faster than
the duration of the LDA selfconsistency cycle if only orbitals for the occupied and first few unoccupied
states are considered.
In contrast, a recent comparative study of several electronic structure codes
estimated that eDMFT takes 5000 times more computer time than meta-GGA with
the modified Becke-Johnson potential~\cite{beyond_dft}.

Finally, in LDA+$U$ the orbital-resolved occupation matrix is used to compute
the potential matrix, which is employed to supplement the Hamiltonian by
correlation effects. This step may be performed as a second variation step,
i.e., one may first compute the LDA eigenvalues and eigenfunctions in first
variation, followed by the calculation of the LDA+$U$ eigenvalues and eigenfunctions
in second variation~\cite{PhysRevB.60.10763}.
Similarly, in the selfconsistent moment method
one may first compute the LDA eigenvalues and eigenfunctions in first variation,
followed by the construction of the spectral function from the first four
moment matrices in second variation. This second variation approach is
expected to take only a fraction of the computing time of the first variation step.

In comparison to LDA+DMFT, one drawback of the selfconsistent moment method
is that it does not give access to the imaginary part of the selfenergy.
However, since the selfconsistent moment method has been used before successfully
to study finite-temperature magnetism~\cite{bcc_iron_Nolting1995,nickel_PhysRevB.40.5015,nickel_Borgiel1990},
we do not expect this to be a major 
drawback for this application. Moreover, many response properties of interest
in spintronics are not strongly dependent on the electron correlation selfenergy
in a large range of realistic materials, 
because they are at most sensitive to the electron lifetime at the Fermi surface,
which is often determined by scattering rather than electron correlation effects.
In such cases, the transfer of spectral weight between states of different
energy might be more important for the accurate computation of the response function
than the imaginary part of the selfenergy.
For example, already simple band shifts due to electron correlation effects
may sometimes improve the agreement of AHE to experiment~\cite{PhysRevB.105.115135}.
Concerning the description of magnetism at finite temperatures
LDA+DMFT studies suggest that missing long wavelength spin waves may lead to an
overestimation of the Curie temperature in some cases~\cite{PhysRevLett.87.067205}. 
We suspect that the selfconsistent moment method suffers from a similar
problem whenever the smallest possible magnetic unit cell is chosen and
when translational invariance is thus enforced.

In comparison to LDA+$U$ the selfconsistent spectral moment method is not only
expected to improve the description of magnetism at finite temperatures, but
additionally it is expected to capture spectral features that arise from the
splitting of bands into lower and upper Hubbard bands, similarly to LDA+DMFT.
For example, it has
been shown~\cite{nickel_PhysRevB.40.5015} to reproduce the valence band
satellite in Ni.

\section{Summary}
\label{sec_summary}
We show that for a general crystal lattice Hamiltonian which  describes
$N_{\rm W}$ electronic orbitals and which includes the Coulomb
interaction the spectral function may be found
within an approximation that assumes that the electronic
structure is given in terms of $2N_{\rm W}$ bands
and $2N_{\rm W}$ spectral weight factors -- $N_{\rm W}$ band energies and
spectral weights describe the lower
Hubbard bands, while $N_{\rm W}$ band energies and spectral weights describe
the upper
Hubbard bands. For this purpose we generalize the standard
two-pole approximation of the selfconsistent moment method
to the case of many bands.
We argue that the problem of constructing $2N_{\rm W}$ bands
and $2N_{\rm W}$ spectral weights
from four $N_{\rm W}\times N_{\rm W}$ spectral moment matrices may
be considered as a generalization of the well-known problem
of finding the $N_{\rm W}$ eigenstates of an $N_{\rm W}\times N_{\rm W}$
single-particle Hamiltonian matrix of a closed quantum system.
We describe how the
higher correlation functions of the
type $\langle c^{\dagger}_{i\alpha}c^{\dagger}_{j\beta}c_{l\gamma} c_{m\delta}\rangle$
may be obtained consistently within this approach by
employing the state vectors and state energies of the single particle spectral
function.
Moreover, we discuss how the response functions may be computed.
We present applications of this approach to the Hubbard-Rashba model.
Our findings suggest that the many-band spectral-moment method
may replace the standard LDA+$U$ approach to
correlated electrons when the temperature dependence of the electronic
structure or of the response functions needs to be determined.
We propose that the many-band spectral-moment method may also
be used instead of Hartree-Fock in exploratory model calculations
of phase diagrams in the search of new quantum states that arise
from the interplay of SOI and correlation effects.

\section*{Acknowledgments}We acknowledge financial support from
Leibniz Collaborative Excellence project OptiSPIN $-$ Optical Control
of Nanoscale Spin Textures, funding  under SPP 2137 ``Skyrmionics" of
the DFG
and Sino-German research project DISTOMAT (DFG project MO \mbox{1731/10-1}).
We gratefully acknowledge financial support from the European Research
Council (ERC) under
the European Union's Horizon 2020 research and innovation
program (Grant No. 856538, project ``3D MAGiC'').
The work was also supported by the Deutsche Forschungsgemeinschaft
(DFG, German Research Foundation) $-$ TRR 173 $-$ 268565370 (project
A11), TRR 288 $-$ 422213477 (project B06).  We  also gratefully
acknowledge the J\"ulich
Supercomputing Centre and RWTH Aachen University for providing
computational
resources under project No. jiff40.

\appendix
\section{Spectral Moments in the Hubbard-Rashba model}
\label{sec_appendix_hubbard}
In this section we give explicit expressions for the
moments Eq.~\eqref{eq_specmom_expectvalue}
of the single-particle spectral function evaluated
for the Hubbard-Rashba model introduced in Eq.~\eqref{eq_rashba_hubbard}.
The zeroth moment is given by
\bege
\tilde{M}^{(0)}_{\vn{k}s s'}=\frac{1}{N}
\sum_{lj}e^{i\vn{k}\cdot
  (\vn{R}_{l}-\vn{R}_{j})}
\langle[c_{ls},c^{\dagger}_{js'}]_{+} \rangle=\delta_{s s'},
\ee
where $s,s'=\uparrow,\downarrow$.

The spin-diagonal first moments are given by
\bege
\begin{aligned}
&\tilde{M}^{(1)}_{\vn{k}s s}=\frac{1}{N}
\sum_{lj}e^{i\vn{k}\cdot
  (\vn{R}_{l}-\vn{R}_{j})}
\langle[[c_{ls},\mathcal{H}]_{-},c^{\dagger}_{js}]_{+} \rangle\\
&=(\epsilon(\vn{k})-\mu)+U\langle n_{-s} \rangle
\end{aligned}
\ee
while those that couple opposite spins are
\bege
\begin{aligned}
&\tilde{M}^{(1)}_{\vn{k}s, -s}=\frac{1}{N}
\sum_{lj}e^{i\vn{k}\cdot
  (\vn{R}_{l}-\vn{R}_{j})}
\langle[[c_{ls},\mathcal{H}]_{-},c^{\dagger}_{j,-s}]_{+} \rangle\\
&={\rm Re}[\alpha(\vn{k})]-si {\rm Im}[\alpha(\vn{k})]
-\frac{U}{N}\sum_{l}\langle
c^{\dagger}_{l,-s} c_{ls}
\rangle.
\end{aligned}
\ee
Here, we defined
\bege
\alpha(\vn{k})=
\frac{1}{N}
t^{\rm R}
\sum_{<j,l>}e^{i\vn{k}\cdot (\vn{R}_{j}-\vn{R}_{l})}
[d_{x j l}+ i d_{y j l}],
\ee
where $d_{x j l}$
and $ d_{y j l}$
are the $x$ and $y$ components, respectively,
of the distance vector $\vn{d}_{jl}$ used in the
Rashba-type SOI in Eq.~\eqref{eq_rashba_hubbard}.

For equal spins the second moments are
\bege
\begin{aligned}
&\tilde{M}^{(2)}_{\vn{k}s s}=\frac{1}{N}
\sum_{lj}e^{i\vn{k}\cdot
  (\vn{R}_{l}-\vn{R}_{j})}
\langle[
[c_{ls},\mathcal{H}]_{-},
[\mathcal{H},c^{\dagger}_{js}]_{-}
]_{+} \rangle\\
&=(\epsilon(\vn{k})-\mu)^2
+2 U\langle n_{-s} \rangle (\epsilon(\vn{k})-\mu)
+ U^2\langle n_{-s}\rangle\\
&-2 U {\rm Re}
\left[
\alpha(\vn{k})\frac{1}{N}\sum_{l}
\langle
c^{\dagger}_{l\downarrow}c_{l\uparrow}
\rangle
\right]+|\alpha(\vn{k})|^{2},
\end{aligned}
\ee
while they are
\bege
\begin{aligned}
&\tilde{M}^{(2)}_{\vn{k}s -s}=\frac{1}{N}
\sum_{lj}e^{i\vn{k}\cdot
  (\vn{R}_{l}-\vn{R}_{j})}
\langle[
[c_{ls},\mathcal{H}]_{-},[\mathcal{H},c^{\dagger}_{j,-s}]
]_{+} \rangle\\
&=
[2\epsilon(\vn{k})-2\mu+nU][{\rm Re}[\alpha(\vn{k})]-si {\rm Im}[\alpha(\vn{k})]]
\\
&-[U^2+2U(\epsilon(\vn{k})-\mu)]
\frac{1}{N}\sum_{l}
\langle
c^{\dagger}_{l,-s}c_{ls}
\rangle
\end{aligned}
\ee
for opposite spins.

The third moments are
\bege
\begin{aligned}
&\tilde{M}^{(3)}_{\vn{k}s s}=\frac{1}{N}
\sum_{lj}e^{i\vn{k}\cdot
  (\vn{R}_{l}-\vn{R}_{j})}
\langle[
[[c_{ls},\mathcal{H}]_{-},\mathcal{H}]_{-},
[\mathcal{H},c^{\dagger}_{js}]_{-}
]_{+} \rangle\\
&=[\epsilon(\vn{k})-\mu]^3
+3U\langle n_{-s} \rangle [\epsilon(\vn{k})-\mu]^2
-3 U^2 \langle n_{-s} \rangle\mu\\
&+2 U^2 \epsilon(\vn{k}) n_{-s}
+2 U^2 t_{00}n_{-s}
+U^3 n_{-s}
\\
&-U^2\frac{1}{N}
\sum_{lj}e^{i\vn{k}\cdot
  (\vn{R}_{l}-\vn{R}_{j})}t_{lj}
\langle
c^{\dagger}_{l,-s}c^{\dagger}_{j,-s}c_{l,-s}c_{j,-s}
\rangle
\\
&+U^2\frac{1}{N}
\sum_{lj}
t_{lj}
\langle
(2n_{ls}-1)c^{\dagger}_{l,-s}c_{j,-s}
\rangle
\\
&+U^2\frac{1}{N}
\sum_{lj}e^{i\vn{k}\cdot
  (\vn{R}_{l}-\vn{R}_{j})}t_{lj}
\langle
c^{\dagger}_{js}c^{\dagger}_{l,-s}c_{ls}c_{j,-s}
\rangle\\
&+U^2\frac{1}{N}
\sum_{lj}e^{i\vn{k}\cdot
  (\vn{R}_{l}-\vn{R}_{j})}t_{lj}
\langle
c^{\dagger}_{js}c^{\dagger}_{j,-s}c_{ls}c_{l,-s}
\rangle\\
&+U^2\frac{2}{N}\sum_{l}
{\rm Re}\left[
\alpha(\vn{k})
\langle
c^{\dagger}_{l,-s}c_{l,s}
\rangle
\right]
\\
&+3|\alpha(\vn{k})|^2 [\epsilon(\vn{k})-\mu]
+|\alpha(\vn{k})|^2 U (n+n_{-s})\\
&+6 U [\epsilon(\vn{k})-\mu] 
{\rm Re}\left[ \alpha(\vn{k})  \frac{1}{N}\sum_{l}
\langle
c^{\dagger}_{l,-s}c_{ls}
\rangle \right]\\
&+U^2\frac{2}{N}
\sum_{lj}
{\rm Re}\left[
e^{i\vn{k}\cdot
  (\vn{R}_{l}-\vn{R}_{j})}\alpha_{lj}
\langle
c^{\dagger}_{l,-s}c^{\dagger}_{j,-s}c_{ls}c_{j,-s}
\rangle
\right],
\\
\end{aligned}
\ee
when the spins are parallel, and
\bege
\begin{aligned}
&\tilde{M}^{(3)}_{\vn{k}\uparrow \downarrow}=\frac{1}{N}
\sum_{lj}e^{i\vn{k}\cdot
  (\vn{R}_{l}-\vn{R}_{j})}
\langle[
[[c_{l\uparrow},\mathcal{H}]_{-},\mathcal{H}]_{-},[\mathcal{H},c^{\dagger}_{j\downarrow}]
]_{+} \rangle\\
&=-3[\alpha(\vn{k})]^{*}[\epsilon(\vn{k})-\mu]^2
-[\alpha(\vn{k})]^{*}|\alpha(\vn{k})|^2
-U^2 [\alpha(\vn{k})]^{*}  n
\\
&- \left (
 3[\epsilon(\vn{k})-\mu]^2
+2 |\alpha(\vn{k})|^2 
+2U
[\epsilon(\vn{k})-\mu]
\right)
   \frac{U}{N}
\sum_{l}
\langle
c^{\dagger}_{l\downarrow }c_{l\uparrow }
\rangle\\
&+U^2 \frac{1}{N}
\sum_{lj}
t_{lj}
\langle
c^{\dagger}_{l\downarrow }c_{j\uparrow }
\rangle\\
&-U
\left[
\alpha^{*}(\vn{k})
\right]^2
\frac{1}{N}
\sum_{l}
\langle
c^{\dagger}_{l\uparrow }c_{l\downarrow }
\rangle\\
&-3 U [\alpha(\vn{k})]^{*} [\epsilon(\vn{k})-\mu] n\\
&-
\left[
U^3 
-U^2\mu
+2U^2 t_{00}
\right]
\frac{1}{N}
\sum_{l}
\langle
c^{\dagger}_{l\downarrow}
c_{l\uparrow}
\rangle\\
&-U^2\frac{1}{N}
\sum_{lj}e^{i\vn{k}\cdot
  (\vn{R}_{l}-\vn{R}_{j})}t_{lj}
\langle c^{\dagger}_{j\uparrow }c^{\dagger}_{l\downarrow }c_{l\uparrow }c_{j\uparrow }\rangle
\\
&-U^2\frac{1}{N}
\sum_{lj}e^{i\vn{k}\cdot
  (\vn{R}_{l}-\vn{R}_{j})}t_{lj}
\langle c^{\dagger}_{l\downarrow }c^{\dagger}_{j\downarrow }c_{j\uparrow  }c_{l\downarrow }\rangle\\
&+U^2\frac{1}{N}
\sum_{lj}e^{i\vn{k}\cdot
  (\vn{R}_{l}-\vn{R}_{j})}\alpha^{*}_{lj}
\langle c^{\dagger}_{j\uparrow }c^{\dagger}_{l\downarrow } c_{j\uparrow }c_{l\downarrow }\rangle\\
&-U^2\frac{1}{N}
\sum_{lj}e^{i\vn{k}\cdot
  (\vn{R}_{l}-\vn{R}_{j})}\alpha^{*}_{lj}
\langle c^{\dagger}_{j\uparrow }c^{\dagger}_{j\downarrow }c_{l\uparrow }c_{l\downarrow }\rangle\\
&-U^2\frac{1}{N}
\sum_{lj}e^{i\vn{k}\cdot
  (\vn{R}_{l}-\vn{R}_{j})}\alpha_{lj}
\langle c^{\dagger}_{l\downarrow }c^{\dagger}_{j\downarrow }c_{l\uparrow }c_{j\uparrow }\rangle\\
&+U^2\frac{1}{N}
\sum_{lj}
t_{lj}
\langle c^{\dagger}_{l\uparrow }c^{\dagger}_{l\downarrow }c_{l\uparrow }c_{j\uparrow }\rangle\\
&-U^2\frac{1}{N}
\sum_{lj}
t_{lj}
\langle c^{\dagger}_{l\downarrow }c^{\dagger}_{j\downarrow }c_{l\uparrow }c_{l\downarrow }\rangle\\
\end{aligned}
\ee
when they are opposite.
Here, we defined
\bege
\alpha_{jl}=
[d_{x j l}+ i d_{y j l}].
\ee

The correlation functions of the
type $\langle c^{\dagger}_{i\alpha}  c^{\dagger}_{j\beta} c_{l\gamma} c_{m\delta} \rangle$
in the expressions of the third moments may be obtained
from the spectral theorem in Eq.~\eqref{eq_corr4_specfun}
and Fourier transformation.

\section{Correlation functions in the Hubbard-Rashba model}
\label{sec_appendix_correlators}
In this appendix we provide examples for explicit expressions
of the
moments $\tilde{K}^{(0)}_{l\gamma m\delta j\beta \vn{k}\alpha}$
and $\tilde{K}^{(1)}_{l\gamma m\delta j\beta \vn{k}\alpha}$
used in Sec.~\ref{sec_corr_fun}.
The zeroth moment is explicitly given by
\bege
\begin{aligned}
  &\tilde{K}^{(0)}_{l\gamma m\delta j\beta \vn{k}\alpha}=\langle[c^{\dagger}_{j\beta} c_{l\gamma} c_{m\delta},c^{\dagger}_{\vn{k}\alpha}]_{+} \rangle\\
  &=
  \frac{1}{\sqrt{N}}
\left[
  \langle
c^{\dagger}_{j\beta}c_{l\gamma}
\rangle
e^{-i\vn{k}\cdot \vn{R}_{m}}\delta_{\alpha\delta}
-
\langle
c^{\dagger}_{j\beta}c_{m\delta}
\rangle
e^{-i\vn{k}\cdot \vn{R}_{l}}\delta_{\alpha\gamma}
\right].
\end{aligned}  
\ee
The thermal averages $\langle c^{\dagger}_{j\beta}c_{l\gamma}\rangle$
and $\langle c^{\dagger}_{j\beta}c_{m\delta} \rangle$
in this expression may be obtained from the spectral theorem Eq.~\eqref{eq_spec_theo_single_part}.

The explicit expression for the first moment
\bege
\tilde{K}^{(1)}_{l\gamma m\delta j\beta \vn{k}\alpha}=\langle[c^{\dagger}_{j\beta} c_{l\gamma} c_{m\delta},[\mathcal{H},c^{\dagger}_{\vn{k}\alpha}]_{-}]_{+} \rangle
\ee
depends on the Hamiltonian $H$.
We list some examples for $\tilde{K}^{(1)}_{l\gamma m\delta j\beta \vn{k}\alpha}$
that we obtain for the Hubbard-Rashba model discussed in Sec.~\ref{sec_rashba_hubbard}.
From
\bege\label{eq_corr4_udud1002}
\begin{aligned}
&\langle[c^{\dagger}_{i\downarrow} c_{i\uparrow} c_{j\downarrow},[H,c^{\dagger}_{l\uparrow}]_{-}]_{+} \rangle=\\
  &-t_{il}\langle c^{\dagger}_{i\downarrow} c_{j\downarrow}  \rangle
  +\alpha_{jl}\langle c^{\dagger}_{i\downarrow} c_{i\uparrow}  \rangle
  -U \delta_{il}\delta_{lj}\langle c^{\dagger}_{l\downarrow} c_{l\downarrow}  \rangle\\
 &+U\delta_{il}\langle c^{\dagger}_{l\uparrow}c^{\dagger}_{l\downarrow} c_{l\uparrow}c_{j\downarrow} \rangle
  +U\delta_{lj}\langle c^{\dagger}_{i\downarrow}c^{\dagger}_{l\uparrow} c_{i\uparrow}c_{l\downarrow} \rangle
\end{aligned}  
\ee
we obtain $\tilde{K}^{(1)}_{i\uparrow j\downarrow i\downarrow \vn{k}\uparrow}$
by performing a Fourier transformation from the site $\vn{R}_{l}$ to the $\vn{k}$-point $\vn{k}$.
As explained in Sec.~\ref{sec_corr_fun}
this moment is required in order to compute the thermal
average $\langle c^{\dagger}_{l\uparrow}  c^{\dagger}_{i\downarrow} c_{i\uparrow} c_{j\downarrow} \rangle$.
However, according to Eq.~\eqref{eq_corr4_udud1002}
additional correlation functions are needed,
namely $\langle c^{\dagger}_{l\uparrow}c^{\dagger}_{l\downarrow} c_{l\uparrow}c_{j\downarrow} \rangle$
and $\langle c^{\dagger}_{i\downarrow}c^{\dagger}_{l\uparrow} c_{i\uparrow}c_{l\downarrow} \rangle$.
As explained in Sec.~\ref{sec_corr_fun} we therefore
use a self-consistent procedure that determines all necessary
correlation functions $\langle c^{\dagger}_{i\alpha}c^{\dagger}_{j\beta}c_{l\gamma} c_{m\delta}\rangle$.

The correlation
function $\langle c^{\dagger}_{l\uparrow}c^{\dagger}_{i\downarrow} c_{j\uparrow}c_{i\uparrow}  \rangle$
vanishes without SOI when the spin-quantization axis is
chosen to be along the $z$-direction, but it may be non-zero
in general when SOI is present. We obtain the corresponding first moment
$\tilde{K}^{(1)}_{j\uparrow i\uparrow i\downarrow \vn{k}\uparrow}$
from
\bege
\begin{aligned}
  &\langle[c^{\dagger}_{i\downarrow} c_{j\uparrow} c_{i\uparrow},[H,c^{\dagger}_{l\uparrow}]_{-}]_{+} \rangle=t_{il}\langle c^{\dagger}_{i\downarrow} c_{j\uparrow}  \rangle-t_{jl}\langle c^{\dagger}_{i\downarrow} c_{i\uparrow}  \rangle\\
  &+U\delta_{jl}
  \langle c^{\dagger}_{i\downarrow}c^{\dagger}_{l\downarrow} c_{i\uparrow}c_{l\downarrow}  \rangle
-U\delta_{il}
  \langle c^{\dagger}_{l\uparrow}c^{\dagger}_{l\downarrow} c_{l\uparrow}c_{j\uparrow}  \rangle  
\end{aligned}
\ee
by performing a Fourier transformation from the site 
$\vn{R}_{l}$ to the $\vn{k}$-point $\vn{k}$.

\section{First moments for response functions}
\label{app_ahe_moments}

In the following we give a few examples for
explicit expressions of the first moment of the
commutator spectral function obtained
by evaluating Eq.~\eqref{eq_moment_D1}
for the Hubbard-Rashba model:
\bege
\begin{aligned}
  &\tilde{D}^{(1)}_{\vn{k}\uparrow\uparrow\uparrow\uparrow}=
        [\alpha(\vn{k})]^{*}\langle c^{\dagger}_{0\downarrow} c_{0\uparrow}  \rangle
        + \alpha(\vn{k})\langle c^{\dagger}_{0\uparrow} c_{0\downarrow}  \rangle\\
        &+U \sum_{j} e^{-i \vn{k}\cdot [\vn{R}_{j}-\vn{R}_{0}]}
        \langle c^{\dagger}_{0\uparrow} c^{\dagger}_{0\downarrow} c_{j\uparrow} c_{0\downarrow}   \rangle\\
        &+U \sum_{j} e^{i \vn{k}\cdot [\vn{R}_{j}-\vn{R}_{0}]}
        \langle c^{\dagger}_{j\uparrow} c^{\dagger}_{0\downarrow} c_{0\uparrow} c_{0\downarrow}   \rangle\\
        &-2\frac{U}{N} \sum_{j,l} e^{i \vn{k}\cdot [\vn{R}_{j}-\vn{R}_{l}]}
        \langle c^{\dagger}_{j\uparrow} c^{\dagger}_{0\downarrow} c_{l\uparrow} c_{0\downarrow}   \rangle        
\end{aligned}  
\ee
and
\bege
\begin{aligned}
  &\tilde{D}^{(1)}_{\vn{k}\uparrow\downarrow\downarrow\uparrow}=
  2 \alpha(\vn{k})\langle c^{\dagger}_{0\downarrow} c_{0\uparrow}  \rangle\\
      &+U \sum_{j} e^{-i \vn{k}\cdot [\vn{R}_{j}-\vn{R}_{0}]}
       \langle c^{\dagger}_{0\uparrow} c^{\dagger}_{0\downarrow} c_{j\uparrow} c_{0\downarrow}   \rangle\\
       &+\frac{U}{N} \sum_{j,l} e^{i \vn{k}\cdot [\vn{R}_{j}-\vn{R}_{l}]}
       \langle c^{\dagger}_{j\uparrow} c^{\dagger}_{0\uparrow} c_{0\uparrow} c_{l\uparrow}   \rangle\\
       &+U \sum_{j} e^{i \vn{k}\cdot [\vn{R}_{j}-\vn{R}_{0}]}
       \langle c^{\dagger}_{0\uparrow} c^{\dagger}_{j\downarrow} c_{0\uparrow} c_{0\downarrow}   \rangle\\
       &-\frac{U}{N} \sum_{j,l} e^{i \vn{k}\cdot [-\vn{R}_{j}-\vn{R}_{l}]}
       \langle c^{\dagger}_{0\uparrow} c^{\dagger}_{0\uparrow} c_{j\uparrow} c_{l\uparrow}   \rangle\\
&-\frac{U}{N} \sum_{j,l} e^{i \vn{k}\cdot [\vn{R}_{j}-\vn{R}_{l}]}
\langle c^{\dagger}_{0\downarrow} c^{\dagger}_{j\downarrow} c_{0\downarrow} c_{l\downarrow}   \rangle\\
&-\frac{U}{N} \sum_{j,l} e^{i \vn{k}\cdot [\vn{R}_{j}+\vn{R}_{l}]}
\langle c^{\dagger}_{j\downarrow} c^{\dagger}_{l\downarrow} c_{0\downarrow} c_{0\downarrow}   \rangle.\\
\end{aligned}  
\ee  
The correlation functions of the
type $\langle c^{\dagger}_{i\alpha}  c^{\dagger}_{j\beta} c_{l\gamma} c_{m\delta} \rangle$
in these expressions may be obtained
from the spectral theorem in Eq.~\eqref{eq_corr4_specfun}
and Fourier transformation.

\bibliography{spectral}

\begin{thebibliography}{44}%
\makeatletter
\providecommand \@ifxundefined [1]{%
 \@ifx{#1\undefined}
}%
\providecommand \@ifnum [1]{%
 \ifnum #1\expandafter \@firstoftwo
 \else \expandafter \@secondoftwo
 \fi
}%
\providecommand \@ifx [1]{%
 \ifx #1\expandafter \@firstoftwo
 \else \expandafter \@secondoftwo
 \fi
}%
\providecommand \natexlab [1]{#1}%
\providecommand \enquote  [1]{``#1''}%
\providecommand \bibnamefont  [1]{#1}%
\providecommand \bibfnamefont [1]{#1}%
\providecommand \citenamefont [1]{#1}%
\providecommand \href@noop [0]{\@secondoftwo}%
\providecommand \href [0]{\begingroup \@sanitize@url \@href}%
\providecommand \@href[1]{\@@startlink{#1}\@@href}%
\providecommand \@@href[1]{\endgroup#1\@@endlink}%
\providecommand \@sanitize@url [0]{\catcode `\\12\catcode `\$12\catcode
  `\&12\catcode `\#12\catcode `\^12\catcode `\_12\catcode `\%12\relax}%
\providecommand \@@startlink[1]{}%
\providecommand \@@endlink[0]{}%
\providecommand \url  [0]{\begingroup\@sanitize@url \@url }%
\providecommand \@url [1]{\endgroup\@href {#1}{\urlprefix }}%
\providecommand \urlprefix  [0]{URL }%
\providecommand \Eprint [0]{\href }%
\providecommand \doibase [0]{https://doi.org/}%
\providecommand \selectlanguage [0]{\@gobble}%
\providecommand \bibinfo  [0]{\@secondoftwo}%
\providecommand \bibfield  [0]{\@secondoftwo}%
\providecommand \translation [1]{[#1]}%
\providecommand \BibitemOpen [0]{}%
\providecommand \bibitemStop [0]{}%
\providecommand \bibitemNoStop [0]{.\EOS\space}%
\providecommand \EOS [0]{\spacefactor3000\relax}%
\providecommand \BibitemShut  [1]{\csname bibitem#1\endcsname}%
\let\auto@bib@innerbib\@empty
\bibitem [{\citenamefont {Himmetoglu}\ \emph {et~al.}(2014)\citenamefont
  {Himmetoglu}, \citenamefont {Floris}, \citenamefont {de~Gironcoli},\ and\
  \citenamefont {Cococcioni}}]{review_ldau}%
  \BibitemOpen
  \bibfield  {author} {\bibinfo {author} {\bibfnamefont {B.}~\bibnamefont
  {Himmetoglu}}, \bibinfo {author} {\bibfnamefont {A.}~\bibnamefont {Floris}},
  \bibinfo {author} {\bibfnamefont {S.}~\bibnamefont {de~Gironcoli}},\ and\
  \bibinfo {author} {\bibfnamefont {M.}~\bibnamefont {Cococcioni}},\ }\bibfield
   {title} {\bibinfo {title} {Hubbard-corrected dft energy functionals: The
  lda+u description of correlated systems},\ }\href
  {https://doi.org/https://doi.org/10.1002/qua.24521} {\bibfield  {journal}
  {\bibinfo  {journal} {International Journal of Quantum Chemistry}\ }\textbf
  {\bibinfo {volume} {114}},\ \bibinfo {pages} {14} (\bibinfo {year}
  {2014})}\BibitemShut {NoStop}%
\bibitem [{\citenamefont {Anisimov}\ \emph {et~al.}(1991)\citenamefont
  {Anisimov}, \citenamefont {Zaanen},\ and\ \citenamefont
  {Andersen}}]{PhysRevB.44.943}%
  \BibitemOpen
  \bibfield  {author} {\bibinfo {author} {\bibfnamefont {V.~I.}\ \bibnamefont
  {Anisimov}}, \bibinfo {author} {\bibfnamefont {J.}~\bibnamefont {Zaanen}},\
  and\ \bibinfo {author} {\bibfnamefont {O.~K.}\ \bibnamefont {Andersen}},\
  }\bibfield  {title} {\bibinfo {title} {Band theory and mott insulators:
  Hubbard u instead of stoner i},\ }\href
  {https://doi.org/10.1103/PhysRevB.44.943} {\bibfield  {journal} {\bibinfo
  {journal} {Phys. Rev. B}\ }\textbf {\bibinfo {volume} {44}},\ \bibinfo
  {pages} {943} (\bibinfo {year} {1991})}\BibitemShut {NoStop}%
\bibitem [{\citenamefont {Anisimov}\ \emph {et~al.}(1993)\citenamefont
  {Anisimov}, \citenamefont {Solovyev}, \citenamefont {Korotin}, \citenamefont
  {Czy\ifmmode~\dot{z}\else \.{z}\fi{}yk},\ and\ \citenamefont
  {Sawatzky}}]{PhysRevB.48.16929}%
  \BibitemOpen
  \bibfield  {author} {\bibinfo {author} {\bibfnamefont {V.~I.}\ \bibnamefont
  {Anisimov}}, \bibinfo {author} {\bibfnamefont {I.~V.}\ \bibnamefont
  {Solovyev}}, \bibinfo {author} {\bibfnamefont {M.~A.}\ \bibnamefont
  {Korotin}}, \bibinfo {author} {\bibfnamefont {M.~T.}\ \bibnamefont
  {Czy\ifmmode~\dot{z}\else \.{z}\fi{}yk}},\ and\ \bibinfo {author}
  {\bibfnamefont {G.~A.}\ \bibnamefont {Sawatzky}},\ }\bibfield  {title}
  {\bibinfo {title} {Density-functional theory and nio photoemission spectra},\
  }\href {https://doi.org/10.1103/PhysRevB.48.16929} {\bibfield  {journal}
  {\bibinfo  {journal} {Phys. Rev. B}\ }\textbf {\bibinfo {volume} {48}},\
  \bibinfo {pages} {16929} (\bibinfo {year} {1993})}\BibitemShut {NoStop}%
\bibitem [{\citenamefont {Solovyev}\ \emph {et~al.}(1994)\citenamefont
  {Solovyev}, \citenamefont {Dederichs},\ and\ \citenamefont
  {Anisimov}}]{PhysRevB.50.16861}%
  \BibitemOpen
  \bibfield  {author} {\bibinfo {author} {\bibfnamefont {I.~V.}\ \bibnamefont
  {Solovyev}}, \bibinfo {author} {\bibfnamefont {P.~H.}\ \bibnamefont
  {Dederichs}},\ and\ \bibinfo {author} {\bibfnamefont {V.~I.}\ \bibnamefont
  {Anisimov}},\ }\bibfield  {title} {\bibinfo {title} {Corrected atomic limit
  in the local-density approximation and the electronic structure of d
  impurities in rb},\ }\href {https://doi.org/10.1103/PhysRevB.50.16861}
  {\bibfield  {journal} {\bibinfo  {journal} {Phys. Rev. B}\ }\textbf {\bibinfo
  {volume} {50}},\ \bibinfo {pages} {16861} (\bibinfo {year}
  {1994})}\BibitemShut {NoStop}%
\bibitem [{\citenamefont {Lichtenstein}\ \emph {et~al.}(2001)\citenamefont
  {Lichtenstein}, \citenamefont {Katsnelson},\ and\ \citenamefont
  {Kotliar}}]{PhysRevLett.87.067205}%
  \BibitemOpen
  \bibfield  {author} {\bibinfo {author} {\bibfnamefont {A.~I.}\ \bibnamefont
  {Lichtenstein}}, \bibinfo {author} {\bibfnamefont {M.~I.}\ \bibnamefont
  {Katsnelson}},\ and\ \bibinfo {author} {\bibfnamefont {G.}~\bibnamefont
  {Kotliar}},\ }\bibfield  {title} {\bibinfo {title} {Finite-temperature
  magnetism of transition metals: An ab initio dynamical mean-field theory},\
  }\href {https://doi.org/10.1103/PhysRevLett.87.067205} {\bibfield  {journal}
  {\bibinfo  {journal} {Phys. Rev. Lett.}\ }\textbf {\bibinfo {volume} {87}},\
  \bibinfo {pages} {067205} (\bibinfo {year} {2001})}\BibitemShut {NoStop}%
\bibitem [{\citenamefont {Schlotter}\ \emph {et~al.}(2018)\citenamefont
  {Schlotter}, \citenamefont {Agrawal},\ and\ \citenamefont
  {Beach}}]{doi:10.1063/1.5038353}%
  \BibitemOpen
  \bibfield  {author} {\bibinfo {author} {\bibfnamefont {S.}~\bibnamefont
  {Schlotter}}, \bibinfo {author} {\bibfnamefont {P.}~\bibnamefont {Agrawal}},\
  and\ \bibinfo {author} {\bibfnamefont {G.~S.~D.}\ \bibnamefont {Beach}},\
  }\bibfield  {title} {\bibinfo {title} {Temperature dependence of the
  dzyaloshinskii-moriya interaction in pt/co/cu thin film heterostructures},\
  }\href {https://doi.org/10.1063/1.5038353} {\bibfield  {journal} {\bibinfo
  {journal} {Applied Physics Letters}\ }\textbf {\bibinfo {volume} {113}},\
  \bibinfo {pages} {092402} (\bibinfo {year} {2018})}\BibitemShut {NoStop}%
\bibitem [{\citenamefont {Kim}\ \emph {et~al.}(2014)\citenamefont {Kim},
  \citenamefont {Sinha}, \citenamefont {Mitani}, \citenamefont {Hayashi},
  \citenamefont {Takahashi}, \citenamefont {Maekawa}, \citenamefont
  {Yamanouchi},\ and\ \citenamefont {Ohno}}]{PhysRevB.89.174424}%
  \BibitemOpen
  \bibfield  {author} {\bibinfo {author} {\bibfnamefont {J.}~\bibnamefont
  {Kim}}, \bibinfo {author} {\bibfnamefont {J.}~\bibnamefont {Sinha}}, \bibinfo
  {author} {\bibfnamefont {S.}~\bibnamefont {Mitani}}, \bibinfo {author}
  {\bibfnamefont {M.}~\bibnamefont {Hayashi}}, \bibinfo {author} {\bibfnamefont
  {S.}~\bibnamefont {Takahashi}}, \bibinfo {author} {\bibfnamefont
  {S.}~\bibnamefont {Maekawa}}, \bibinfo {author} {\bibfnamefont
  {M.}~\bibnamefont {Yamanouchi}},\ and\ \bibinfo {author} {\bibfnamefont
  {H.}~\bibnamefont {Ohno}},\ }\bibfield  {title} {\bibinfo {title} {Anomalous
  temperature dependence of current-induced torques in
  $\text{CoFeB}/\text{MgO}$ heterostructures with ta-based underlayers},\
  }\href {https://doi.org/10.1103/PhysRevB.89.174424} {\bibfield  {journal}
  {\bibinfo  {journal} {Phys. Rev. B}\ }\textbf {\bibinfo {volume} {89}},\
  \bibinfo {pages} {174424} (\bibinfo {year} {2014})}\BibitemShut {NoStop}%
\bibitem [{\citenamefont {Qiu}\ \emph {et~al.}(2014)\citenamefont {Qiu},
  \citenamefont {Deorani}, \citenamefont {Narayanapillai}, \citenamefont {Lee},
  \citenamefont {Lee}, \citenamefont {Lee},\ and\ \citenamefont
  {Yang}}]{angular_and_temperature_dependence_Ta_CoFeB_MgO}%
  \BibitemOpen
  \bibfield  {author} {\bibinfo {author} {\bibfnamefont {X.}~\bibnamefont
  {Qiu}}, \bibinfo {author} {\bibfnamefont {P.}~\bibnamefont {Deorani}},
  \bibinfo {author} {\bibfnamefont {K.}~\bibnamefont {Narayanapillai}},
  \bibinfo {author} {\bibfnamefont {K.-S.}\ \bibnamefont {Lee}}, \bibinfo
  {author} {\bibfnamefont {K.-J.}\ \bibnamefont {Lee}}, \bibinfo {author}
  {\bibfnamefont {H.-W.}\ \bibnamefont {Lee}},\ and\ \bibinfo {author}
  {\bibfnamefont {H.}~\bibnamefont {Yang}},\ }\bibfield  {title} {\bibinfo
  {title} {Angular and temperature dependence of current induced spin-orbit
  effective fields in ta/cofeb/mgo nanowires},\ }\bibfield  {journal} {\bibinfo
   {journal} {SCIENTIFIC REPORTS}\ }\textbf {\bibinfo {volume} {4}},\ \href
  {https://doi.org/10.1038/srep04491} {10.1038/srep04491} (\bibinfo {year}
  {2014})\BibitemShut {NoStop}%
\bibitem [{\citenamefont {Freimuth}\ \emph {et~al.}(2021)\citenamefont
  {Freimuth}, \citenamefont {Bl\"ugel},\ and\ \citenamefont
  {Mokrousov}}]{magnonicsot}%
  \BibitemOpen
  \bibfield  {author} {\bibinfo {author} {\bibfnamefont {F.}~\bibnamefont
  {Freimuth}}, \bibinfo {author} {\bibfnamefont {S.}~\bibnamefont {Bl\"ugel}},\
  and\ \bibinfo {author} {\bibfnamefont {Y.}~\bibnamefont {Mokrousov}},\
  }\bibfield  {title} {\bibinfo {title} {Effect of magnons on the temperature
  dependence and anisotropy of spin-orbit torque},\ }\href
  {https://doi.org/10.1103/PhysRevB.104.094434} {\bibfield  {journal} {\bibinfo
   {journal} {Phys. Rev. B}\ }\textbf {\bibinfo {volume} {104}},\ \bibinfo
  {pages} {094434} (\bibinfo {year} {2021})}\BibitemShut {NoStop}%
\bibitem [{\citenamefont {Manchon}\ \emph {et~al.}(2019)\citenamefont
  {Manchon}, \citenamefont {\ifmmode~\check{Z}\else \v{Z}\fi{}elezn\'y},
  \citenamefont {Miron}, \citenamefont {Jungwirth}, \citenamefont {Sinova},
  \citenamefont {Thiaville}, \citenamefont {Garello},\ and\ \citenamefont
  {Gambardella}}]{rmp_sot}%
  \BibitemOpen
  \bibfield  {author} {\bibinfo {author} {\bibfnamefont {A.}~\bibnamefont
  {Manchon}}, \bibinfo {author} {\bibfnamefont {J.}~\bibnamefont
  {\ifmmode~\check{Z}\else \v{Z}\fi{}elezn\'y}}, \bibinfo {author}
  {\bibfnamefont {I.~M.}\ \bibnamefont {Miron}}, \bibinfo {author}
  {\bibfnamefont {T.}~\bibnamefont {Jungwirth}}, \bibinfo {author}
  {\bibfnamefont {J.}~\bibnamefont {Sinova}}, \bibinfo {author} {\bibfnamefont
  {A.}~\bibnamefont {Thiaville}}, \bibinfo {author} {\bibfnamefont
  {K.}~\bibnamefont {Garello}},\ and\ \bibinfo {author} {\bibfnamefont
  {P.}~\bibnamefont {Gambardella}},\ }\bibfield  {title} {\bibinfo {title}
  {Current-induced spin-orbit torques in ferromagnetic and antiferromagnetic
  systems},\ }\href {https://doi.org/10.1103/RevModPhys.91.035004} {\bibfield
  {journal} {\bibinfo  {journal} {Rev. Mod. Phys.}\ }\textbf {\bibinfo {volume}
  {91}},\ \bibinfo {pages} {035004} (\bibinfo {year} {2019})}\BibitemShut
  {NoStop}%
\bibitem [{\citenamefont {Wang}\ \emph {et~al.}(2020)\citenamefont {Wang},
  \citenamefont {Ding}, \citenamefont {Moritz}, \citenamefont {Huang},\ and\
  \citenamefont {Devereaux}}]{dc_hall_Hubbard_Wang2020}%
  \BibitemOpen
  \bibfield  {author} {\bibinfo {author} {\bibfnamefont {W.~O.}\ \bibnamefont
  {Wang}}, \bibinfo {author} {\bibfnamefont {J.~K.}\ \bibnamefont {Ding}},
  \bibinfo {author} {\bibfnamefont {B.}~\bibnamefont {Moritz}}, \bibinfo
  {author} {\bibfnamefont {E.~W.}\ \bibnamefont {Huang}},\ and\ \bibinfo
  {author} {\bibfnamefont {T.~P.}\ \bibnamefont {Devereaux}},\ }\bibfield
  {title} {\bibinfo {title} {Dc hall coefficient of the strongly correlated
  hubbard model},\ }\href {https://doi.org/10.1038/s41535-020-00254-w}
  {\bibfield  {journal} {\bibinfo  {journal} {npj Quantum Materials}\ }\textbf
  {\bibinfo {volume} {5}},\ \bibinfo {pages} {51} (\bibinfo {year}
  {2020})}\BibitemShut {NoStop}%
\bibitem [{\citenamefont {Wang}\ \emph {et~al.}(2021)\citenamefont {Wang},
  \citenamefont {Ding}, \citenamefont {Moritz}, \citenamefont {Schattner},
  \citenamefont {Huang},\ and\ \citenamefont
  {Devereaux}}]{PhysRevResearch.3.033033}%
  \BibitemOpen
  \bibfield  {author} {\bibinfo {author} {\bibfnamefont {W.~O.}\ \bibnamefont
  {Wang}}, \bibinfo {author} {\bibfnamefont {J.~K.}\ \bibnamefont {Ding}},
  \bibinfo {author} {\bibfnamefont {B.}~\bibnamefont {Moritz}}, \bibinfo
  {author} {\bibfnamefont {Y.}~\bibnamefont {Schattner}}, \bibinfo {author}
  {\bibfnamefont {E.~W.}\ \bibnamefont {Huang}},\ and\ \bibinfo {author}
  {\bibfnamefont {T.~P.}\ \bibnamefont {Devereaux}},\ }\bibfield  {title}
  {\bibinfo {title} {Numerical approaches for calculating the low-field dc hall
  coefficient of the doped hubbard model},\ }\href
  {https://doi.org/10.1103/PhysRevResearch.3.033033} {\bibfield  {journal}
  {\bibinfo  {journal} {Phys. Rev. Research}\ }\textbf {\bibinfo {volume}
  {3}},\ \bibinfo {pages} {033033} (\bibinfo {year} {2021})}\BibitemShut
  {NoStop}%
\bibitem [{\citenamefont {Auerbach}(2019)}]{PhysRevB.99.115115}%
  \BibitemOpen
  \bibfield  {author} {\bibinfo {author} {\bibfnamefont {A.}~\bibnamefont
  {Auerbach}},\ }\bibfield  {title} {\bibinfo {title} {Equilibrium formulae for
  transverse magnetotransport of strongly correlated metals},\ }\href
  {https://doi.org/10.1103/PhysRevB.99.115115} {\bibfield  {journal} {\bibinfo
  {journal} {Phys. Rev. B}\ }\textbf {\bibinfo {volume} {99}},\ \bibinfo
  {pages} {115115} (\bibinfo {year} {2019})}\BibitemShut {NoStop}%
\bibitem [{\citenamefont {Auerbach}(2018)}]{PhysRevLett.121.066601}%
  \BibitemOpen
  \bibfield  {author} {\bibinfo {author} {\bibfnamefont {A.}~\bibnamefont
  {Auerbach}},\ }\bibfield  {title} {\bibinfo {title} {Hall number of strongly
  correlated metals},\ }\href {https://doi.org/10.1103/PhysRevLett.121.066601}
  {\bibfield  {journal} {\bibinfo  {journal} {Phys. Rev. Lett.}\ }\textbf
  {\bibinfo {volume} {121}},\ \bibinfo {pages} {066601} (\bibinfo {year}
  {2018})}\BibitemShut {NoStop}%
\bibitem [{\citenamefont {R\'ozsa}\ \emph {et~al.}(2017)\citenamefont
  {R\'ozsa}, \citenamefont {Atxitia},\ and\ \citenamefont
  {Nowak}}]{PhysRevB.96.094436}%
  \BibitemOpen
  \bibfield  {author} {\bibinfo {author} {\bibfnamefont {L.}~\bibnamefont
  {R\'ozsa}}, \bibinfo {author} {\bibfnamefont {U.}~\bibnamefont {Atxitia}},\
  and\ \bibinfo {author} {\bibfnamefont {U.}~\bibnamefont {Nowak}},\ }\bibfield
   {title} {\bibinfo {title} {Temperature scaling of the dzyaloshinsky-moriya
  interaction in the spin wave spectrum},\ }\href
  {https://doi.org/10.1103/PhysRevB.96.094436} {\bibfield  {journal} {\bibinfo
  {journal} {Phys. Rev. B}\ }\textbf {\bibinfo {volume} {96}},\ \bibinfo
  {pages} {094436} (\bibinfo {year} {2017})}\BibitemShut {NoStop}%
\bibitem [{\citenamefont {Callen}\ and\ \citenamefont
  {Callen}(1966)}]{CALLEN19661271}%
  \BibitemOpen
  \bibfield  {author} {\bibinfo {author} {\bibfnamefont {H.}~\bibnamefont
  {Callen}}\ and\ \bibinfo {author} {\bibfnamefont {E.}~\bibnamefont
  {Callen}},\ }\bibfield  {title} {\bibinfo {title} {The present status of the
  temperature dependence of magnetocrystalline anisotropy, and the l(l+1)2
  power law},\ }\href
  {https://doi.org/https://doi.org/10.1016/0022-3697(66)90012-6} {\bibfield
  {journal} {\bibinfo  {journal} {Journal of Physics and Chemistry of Solids}\
  }\textbf {\bibinfo {volume} {27}},\ \bibinfo {pages} {1271} (\bibinfo {year}
  {1966})}\BibitemShut {NoStop}%
\bibitem [{\citenamefont {Callen}(1963)}]{PhysRev.130.890}%
  \BibitemOpen
  \bibfield  {author} {\bibinfo {author} {\bibfnamefont {H.~B.}\ \bibnamefont
  {Callen}},\ }\bibfield  {title} {\bibinfo {title} {Green function theory of
  ferromagnetism},\ }\href {https://doi.org/10.1103/PhysRev.130.890} {\bibfield
   {journal} {\bibinfo  {journal} {Phys. Rev.}\ }\textbf {\bibinfo {volume}
  {130}},\ \bibinfo {pages} {890} (\bibinfo {year} {1963})}\BibitemShut
  {NoStop}%
\bibitem [{\citenamefont {Yao}\ \emph {et~al.}(2004)\citenamefont {Yao},
  \citenamefont {Kleinman}, \citenamefont {MacDonald}, \citenamefont {Sinova},
  \citenamefont {Jungwirth}, \citenamefont {Wang}, \citenamefont {Wang},\ and\
  \citenamefont {Niu}}]{PhysRevLett.92.037204}%
  \BibitemOpen
  \bibfield  {author} {\bibinfo {author} {\bibfnamefont {Y.}~\bibnamefont
  {Yao}}, \bibinfo {author} {\bibfnamefont {L.}~\bibnamefont {Kleinman}},
  \bibinfo {author} {\bibfnamefont {A.~H.}\ \bibnamefont {MacDonald}}, \bibinfo
  {author} {\bibfnamefont {J.}~\bibnamefont {Sinova}}, \bibinfo {author}
  {\bibfnamefont {T.}~\bibnamefont {Jungwirth}}, \bibinfo {author}
  {\bibfnamefont {D.-s.}\ \bibnamefont {Wang}}, \bibinfo {author}
  {\bibfnamefont {E.}~\bibnamefont {Wang}},\ and\ \bibinfo {author}
  {\bibfnamefont {Q.}~\bibnamefont {Niu}},\ }\bibfield  {title} {\bibinfo
  {title} {First principles calculation of anomalous hall conductivity in
  ferromagnetic bcc fe},\ }\href
  {https://doi.org/10.1103/PhysRevLett.92.037204} {\bibfield  {journal}
  {\bibinfo  {journal} {Phys. Rev. Lett.}\ }\textbf {\bibinfo {volume} {92}},\
  \bibinfo {pages} {037204} (\bibinfo {year} {2004})}\BibitemShut {NoStop}%
\bibitem [{\citenamefont {Nagaosa}\ \emph {et~al.}(2010)\citenamefont
  {Nagaosa}, \citenamefont {Sinova}, \citenamefont {Onoda}, \citenamefont
  {MacDonald},\ and\ \citenamefont {Ong}}]{RevModPhys.82.1539}%
  \BibitemOpen
  \bibfield  {author} {\bibinfo {author} {\bibfnamefont {N.}~\bibnamefont
  {Nagaosa}}, \bibinfo {author} {\bibfnamefont {J.}~\bibnamefont {Sinova}},
  \bibinfo {author} {\bibfnamefont {S.}~\bibnamefont {Onoda}}, \bibinfo
  {author} {\bibfnamefont {A.~H.}\ \bibnamefont {MacDonald}},\ and\ \bibinfo
  {author} {\bibfnamefont {N.~P.}\ \bibnamefont {Ong}},\ }\bibfield  {title}
  {\bibinfo {title} {Anomalous hall effect},\ }\href
  {https://doi.org/10.1103/RevModPhys.82.1539} {\bibfield  {journal} {\bibinfo
  {journal} {Rev. Mod. Phys.}\ }\textbf {\bibinfo {volume} {82}},\ \bibinfo
  {pages} {1539} (\bibinfo {year} {2010})}\BibitemShut {NoStop}%
\bibitem [{\citenamefont {Wang}\ \emph {et~al.}(2006)\citenamefont {Wang},
  \citenamefont {Yates}, \citenamefont {Souza},\ and\ \citenamefont
  {Vanderbilt}}]{PhysRevB.74.195118}%
  \BibitemOpen
  \bibfield  {author} {\bibinfo {author} {\bibfnamefont {X.}~\bibnamefont
  {Wang}}, \bibinfo {author} {\bibfnamefont {J.~R.}\ \bibnamefont {Yates}},
  \bibinfo {author} {\bibfnamefont {I.}~\bibnamefont {Souza}},\ and\ \bibinfo
  {author} {\bibfnamefont {D.}~\bibnamefont {Vanderbilt}},\ }\bibfield  {title}
  {\bibinfo {title} {Ab initio calculation of the anomalous hall conductivity
  by wannier interpolation},\ }\href
  {https://doi.org/10.1103/PhysRevB.74.195118} {\bibfield  {journal} {\bibinfo
  {journal} {Phys. Rev. B}\ }\textbf {\bibinfo {volume} {74}},\ \bibinfo
  {pages} {195118} (\bibinfo {year} {2006})}\BibitemShut {NoStop}%
\bibitem [{\citenamefont {Freimuth}\ \emph {et~al.}(2014)\citenamefont
  {Freimuth}, \citenamefont {Bl\"ugel},\ and\ \citenamefont
  {Mokrousov}}]{ibcsoit}%
  \BibitemOpen
  \bibfield  {author} {\bibinfo {author} {\bibfnamefont {F.}~\bibnamefont
  {Freimuth}}, \bibinfo {author} {\bibfnamefont {S.}~\bibnamefont {Bl\"ugel}},\
  and\ \bibinfo {author} {\bibfnamefont {Y.}~\bibnamefont {Mokrousov}},\
  }\bibfield  {title} {\bibinfo {title} {Spin-orbit torques in co/pt(111) and
  mn/w(001) magnetic bilayers from first principles},\ }\href@noop {}
  {\bibfield  {journal} {\bibinfo  {journal} {Phys. Rev. B}\ }\textbf {\bibinfo
  {volume} {90}},\ \bibinfo {pages} {174423} (\bibinfo {year}
  {2014})}\BibitemShut {NoStop}%
\bibitem [{\citenamefont {Nolting}\ \emph {et~al.}(1995)\citenamefont
  {Nolting}, \citenamefont {Vega},\ and\ \citenamefont
  {Fauster}}]{bcc_iron_Nolting1995}%
  \BibitemOpen
  \bibfield  {author} {\bibinfo {author} {\bibfnamefont {W.}~\bibnamefont
  {Nolting}}, \bibinfo {author} {\bibfnamefont {A.}~\bibnamefont {Vega}},\ and\
  \bibinfo {author} {\bibfnamefont {T.}~\bibnamefont {Fauster}},\ }\bibfield
  {title} {\bibinfo {title} {Electronic quasiparticle structure of
  ferromagnetic bcc iron},\ }\href {https://doi.org/10.1007/BF01313058}
  {\bibfield  {journal} {\bibinfo  {journal} {Zeitschrift f{\"u}r Physik B
  Condensed Matter}\ }\textbf {\bibinfo {volume} {96}},\ \bibinfo {pages} {357}
  (\bibinfo {year} {1995})}\BibitemShut {NoStop}%
\bibitem [{\citenamefont {Nolting}\ \emph {et~al.}(1989)\citenamefont
  {Nolting}, \citenamefont {Borgiel/}, \citenamefont {Dose},\ and\
  \citenamefont {Fauster}}]{nickel_PhysRevB.40.5015}%
  \BibitemOpen
  \bibfield  {author} {\bibinfo {author} {\bibfnamefont {W.}~\bibnamefont
  {Nolting}}, \bibinfo {author} {\bibfnamefont {W.}~\bibnamefont {Borgiel/}},
  \bibinfo {author} {\bibfnamefont {V.}~\bibnamefont {Dose}},\ and\ \bibinfo
  {author} {\bibfnamefont {T.}~\bibnamefont {Fauster}},\ }\bibfield  {title}
  {\bibinfo {title} {Finite-temperature ferromagnetism of nickel},\ }\href
  {https://doi.org/10.1103/PhysRevB.40.5015} {\bibfield  {journal} {\bibinfo
  {journal} {Phys. Rev. B}\ }\textbf {\bibinfo {volume} {40}},\ \bibinfo
  {pages} {5015} (\bibinfo {year} {1989})}\BibitemShut {NoStop}%
\bibitem [{\citenamefont {Borgiel}\ and\ \citenamefont
  {Nolting}(1990)}]{nickel_Borgiel1990}%
  \BibitemOpen
  \bibfield  {author} {\bibinfo {author} {\bibfnamefont {W.}~\bibnamefont
  {Borgiel}}\ and\ \bibinfo {author} {\bibfnamefont {W.}~\bibnamefont
  {Nolting}},\ }\bibfield  {title} {\bibinfo {title} {Many body contributions
  to the electronic structure of nickel},\ }\href
  {https://doi.org/10.1007/BF01307842} {\bibfield  {journal} {\bibinfo
  {journal} {Zeitschrift f{\"u}r Physik B Condensed Matter}\ }\textbf {\bibinfo
  {volume} {78}},\ \bibinfo {pages} {241} (\bibinfo {year} {1990})}\BibitemShut
  {NoStop}%
\bibitem [{\citenamefont {Cr\'epieux}\ and\ \citenamefont
  {Bruno}(2001)}]{crepieux_bruno_ahe}%
  \BibitemOpen
  \bibfield  {author} {\bibinfo {author} {\bibfnamefont {A.}~\bibnamefont
  {Cr\'epieux}}\ and\ \bibinfo {author} {\bibfnamefont {P.}~\bibnamefont
  {Bruno}},\ }\bibfield  {title} {\bibinfo {title} {Theory of the anomalous
  hall effect from the kubo formula and the dirac equation},\ }\href@noop {}
  {\bibfield  {journal} {\bibinfo  {journal} {Phys. Rev. B}\ }\textbf {\bibinfo
  {volume} {64}},\ \bibinfo {pages} {014416} (\bibinfo {year}
  {2001})}\BibitemShut {NoStop}%
\bibitem [{\citenamefont {Bastin}\ \emph {et~al.}(1971)\citenamefont {Bastin},
  \citenamefont {Lewiner}, \citenamefont {Betbeder-Matibet},\ and\
  \citenamefont {Nozi\`eres}}]{bastin_conductivity}%
  \BibitemOpen
  \bibfield  {author} {\bibinfo {author} {\bibfnamefont {A.}~\bibnamefont
  {Bastin}}, \bibinfo {author} {\bibfnamefont {C.}~\bibnamefont {Lewiner}},
  \bibinfo {author} {\bibfnamefont {O.}~\bibnamefont {Betbeder-Matibet}},\ and\
  \bibinfo {author} {\bibfnamefont {P.}~\bibnamefont {Nozi\`eres}},\
  }\href@noop {} {\bibfield  {journal} {\bibinfo  {journal} {J. Phys. Chem.
  Solids}\ }\textbf {\bibinfo {volume} {32}},\ \bibinfo {pages} {1811}
  (\bibinfo {year} {1971})}\BibitemShut {NoStop}%
\bibitem [{\citenamefont
  {Nolting}(1975)}]{hubbard_model_electrical_conductivity}%
  \BibitemOpen
  \bibfield  {author} {\bibinfo {author} {\bibfnamefont {W.}~\bibnamefont
  {Nolting}},\ }\bibfield  {title} {\bibinfo {title} {On the connection between
  electric and magnetic properties of the hubbard model i. electrical
  conductivity},\ }\href
  {https://doi.org/https://doi.org/10.1002/pssb.2220700209} {\bibfield
  {journal} {\bibinfo  {journal} {physica status solidi (b)}\ }\textbf
  {\bibinfo {volume} {70}},\ \bibinfo {pages} {505} (\bibinfo {year}
  {1975})}\BibitemShut {NoStop}%
\bibitem [{\citenamefont {Geipel}\ and\ \citenamefont
  {Nolting}(1988)}]{PhysRevB.38.2608}%
  \BibitemOpen
  \bibfield  {author} {\bibinfo {author} {\bibfnamefont {G.}~\bibnamefont
  {Geipel}}\ and\ \bibinfo {author} {\bibfnamefont {W.}~\bibnamefont
  {Nolting}},\ }\bibfield  {title} {\bibinfo {title} {Ferromagnetism in the
  strongly correlated hubbard model},\ }\href
  {https://doi.org/10.1103/PhysRevB.38.2608} {\bibfield  {journal} {\bibinfo
  {journal} {Phys. Rev. B}\ }\textbf {\bibinfo {volume} {38}},\ \bibinfo
  {pages} {2608} (\bibinfo {year} {1988})}\BibitemShut {NoStop}%
\bibitem [{\citenamefont {Pizzi}\ \emph {et~al.}(2020)\citenamefont {Pizzi},
  \citenamefont {Vitale}, \citenamefont {Arita}, \citenamefont {Bl\"ugel},
  \citenamefont {Freimuth}, \citenamefont {G\'eranton}, \citenamefont
  {Gibertini}, \citenamefont {Gresch}, \citenamefont {Johnson}, \citenamefont
  {Koretsune},\ and\ \citenamefont {et~al.}}]{wannier90communitycode}%
  \BibitemOpen
  \bibfield  {author} {\bibinfo {author} {\bibfnamefont {G.}~\bibnamefont
  {Pizzi}}, \bibinfo {author} {\bibfnamefont {V.}~\bibnamefont {Vitale}},
  \bibinfo {author} {\bibfnamefont {R.}~\bibnamefont {Arita}}, \bibinfo
  {author} {\bibfnamefont {S.}~\bibnamefont {Bl\"ugel}}, \bibinfo {author}
  {\bibfnamefont {F.}~\bibnamefont {Freimuth}}, \bibinfo {author}
  {\bibfnamefont {G.}~\bibnamefont {G\'eranton}}, \bibinfo {author}
  {\bibfnamefont {M.}~\bibnamefont {Gibertini}}, \bibinfo {author}
  {\bibfnamefont {D.}~\bibnamefont {Gresch}}, \bibinfo {author} {\bibfnamefont
  {C.}~\bibnamefont {Johnson}}, \bibinfo {author} {\bibfnamefont
  {T.}~\bibnamefont {Koretsune}},\ and\ \bibinfo {author} {\bibnamefont
  {et~al.}},\ }\bibfield  {title} {\bibinfo {title} {Wannier90 as a community
  code: new features and applications},\ }\href@noop {} {\bibfield  {journal}
  {\bibinfo  {journal} {J. Phys.: Condens. Matter}\ }\textbf {\bibinfo {volume}
  {32}},\ \bibinfo {pages} {165902} (\bibinfo {year} {2020})}\BibitemShut
  {NoStop}%
\bibitem [{\citenamefont {Anisimov}\ \emph {et~al.}(2007)\citenamefont
  {Anisimov}, \citenamefont {Kozhevnikov}, \citenamefont {Korotin},
  \citenamefont {Lukoyanov},\ and\ \citenamefont
  {Khafizullin}}]{ldau_wannier_Anisimov_2007}%
  \BibitemOpen
  \bibfield  {author} {\bibinfo {author} {\bibfnamefont {V.~I.}\ \bibnamefont
  {Anisimov}}, \bibinfo {author} {\bibfnamefont {A.~V.}\ \bibnamefont
  {Kozhevnikov}}, \bibinfo {author} {\bibfnamefont {M.~A.}\ \bibnamefont
  {Korotin}}, \bibinfo {author} {\bibfnamefont {A.~V.}\ \bibnamefont
  {Lukoyanov}},\ and\ \bibinfo {author} {\bibfnamefont {D.~A.}\ \bibnamefont
  {Khafizullin}},\ }\bibfield  {title} {\bibinfo {title} {Orbital density
  functional as a means to restore the discontinuities in the total-energy
  derivative and the exchange{\textendash}correlation potential},\ }\href
  {https://doi.org/10.1088/0953-8984/19/10/106206} {\bibfield  {journal}
  {\bibinfo  {journal} {Journal of Physics: Condensed Matter}\ }\textbf
  {\bibinfo {volume} {19}},\ \bibinfo {pages} {106206} (\bibinfo {year}
  {2007})}\BibitemShut {NoStop}%
\bibitem [{\citenamefont {Witczak-Krempa}\ \emph {et~al.}(2014)\citenamefont
  {Witczak-Krempa}, \citenamefont {Chen}, \citenamefont {Kim},\ and\
  \citenamefont {Balents}}]{corr_plus_soi_balents}%
  \BibitemOpen
  \bibfield  {author} {\bibinfo {author} {\bibfnamefont {W.}~\bibnamefont
  {Witczak-Krempa}}, \bibinfo {author} {\bibfnamefont {G.}~\bibnamefont
  {Chen}}, \bibinfo {author} {\bibfnamefont {Y.~B.}\ \bibnamefont {Kim}},\ and\
  \bibinfo {author} {\bibfnamefont {L.}~\bibnamefont {Balents}},\ }\bibfield
  {title} {\bibinfo {title} {Correlated quantum phenomena in the strong
  spin-orbit regime},\ }\href
  {https://doi.org/10.1146/annurev-conmatphys-020911-125138} {\bibfield
  {journal} {\bibinfo  {journal} {Annual Review of Condensed Matter Physics}\
  }\textbf {\bibinfo {volume} {5}},\ \bibinfo {pages} {57} (\bibinfo {year}
  {2014})},\ \Eprint
  {https://arxiv.org/abs/https://doi.org/10.1146/annurev-conmatphys-020911-125138}
  {https://doi.org/10.1146/annurev-conmatphys-020911-125138} \BibitemShut
  {NoStop}%
\bibitem [{\citenamefont {Bertinshaw}\ \emph {et~al.}(2019)\citenamefont
  {Bertinshaw}, \citenamefont {Kim}, \citenamefont {Khaliullin},\ and\
  \citenamefont {Kim}}]{square_lattice_iridates}%
  \BibitemOpen
  \bibfield  {author} {\bibinfo {author} {\bibfnamefont {J.}~\bibnamefont
  {Bertinshaw}}, \bibinfo {author} {\bibfnamefont {Y.}~\bibnamefont {Kim}},
  \bibinfo {author} {\bibfnamefont {G.}~\bibnamefont {Khaliullin}},\ and\
  \bibinfo {author} {\bibfnamefont {B.}~\bibnamefont {Kim}},\ }\bibfield
  {title} {\bibinfo {title} {Square lattice iridates},\ }\href
  {https://doi.org/10.1146/annurev-conmatphys-031218-013113} {\bibfield
  {journal} {\bibinfo  {journal} {Annual Review of Condensed Matter Physics}\
  }\textbf {\bibinfo {volume} {10}},\ \bibinfo {pages} {315} (\bibinfo {year}
  {2019})},\ \Eprint
  {https://arxiv.org/abs/https://doi.org/10.1146/annurev-conmatphys-031218-013113}
  {https://doi.org/10.1146/annurev-conmatphys-031218-013113} \BibitemShut
  {NoStop}%
\bibitem [{\citenamefont {Witczak-Krempa}\ and\ \citenamefont
  {Kim}(2012)}]{PhysRevB.85.045124}%
  \BibitemOpen
  \bibfield  {author} {\bibinfo {author} {\bibfnamefont {W.}~\bibnamefont
  {Witczak-Krempa}}\ and\ \bibinfo {author} {\bibfnamefont {Y.~B.}\
  \bibnamefont {Kim}},\ }\bibfield  {title} {\bibinfo {title} {Topological and
  magnetic phases of interacting electrons in the pyrochlore iridates},\ }\href
  {https://doi.org/10.1103/PhysRevB.85.045124} {\bibfield  {journal} {\bibinfo
  {journal} {Phys. Rev. B}\ }\textbf {\bibinfo {volume} {85}},\ \bibinfo
  {pages} {045124} (\bibinfo {year} {2012})}\BibitemShut {NoStop}%
\bibitem [{\citenamefont {Witczak-Krempa}\ \emph {et~al.}(2013)\citenamefont
  {Witczak-Krempa}, \citenamefont {Go},\ and\ \citenamefont
  {Kim}}]{PhysRevB.87.155101}%
  \BibitemOpen
  \bibfield  {author} {\bibinfo {author} {\bibfnamefont {W.}~\bibnamefont
  {Witczak-Krempa}}, \bibinfo {author} {\bibfnamefont {A.}~\bibnamefont {Go}},\
  and\ \bibinfo {author} {\bibfnamefont {Y.~B.}\ \bibnamefont {Kim}},\
  }\bibfield  {title} {\bibinfo {title} {Pyrochlore electrons under pressure,
  heat, and field: Shedding light on the iridates},\ }\href
  {https://doi.org/10.1103/PhysRevB.87.155101} {\bibfield  {journal} {\bibinfo
  {journal} {Phys. Rev. B}\ }\textbf {\bibinfo {volume} {87}},\ \bibinfo
  {pages} {155101} (\bibinfo {year} {2013})}\BibitemShut {NoStop}%
\bibitem [{\citenamefont {Kennedy}\ \emph {et~al.}(2022)\citenamefont
  {Kennedy}, \citenamefont {Júnior}, \citenamefont {Costa},\ and\
  \citenamefont {Santos}}]{metal_insulator_rashba_hubbard}%
  \BibitemOpen
  \bibfield  {author} {\bibinfo {author} {\bibfnamefont {W.}~\bibnamefont
  {Kennedy}}, \bibinfo {author} {\bibfnamefont {S.~d. A.~S.}\ \bibnamefont
  {Júnior}}, \bibinfo {author} {\bibfnamefont {N.~C.}\ \bibnamefont {Costa}},\
  and\ \bibinfo {author} {\bibfnamefont {R.~R.~d.}\ \bibnamefont {Santos}},\
  }\href {https://doi.org/10.48550/ARXIV.2205.08651} {\bibinfo {title}
  {Magnetism and metal-insulator transitions in the rashba-hubbard model}}
  (\bibinfo {year} {2022}),\ \Eprint {https://arxiv.org/abs/2205.08651}
  {arXiv:2205.08651} \BibitemShut {NoStop}%
\bibitem [{\citenamefont {Bercioux}\ and\ \citenamefont
  {Lucignano}(2015)}]{Bercioux_2015}%
  \BibitemOpen
  \bibfield  {author} {\bibinfo {author} {\bibfnamefont {D.}~\bibnamefont
  {Bercioux}}\ and\ \bibinfo {author} {\bibfnamefont {P.}~\bibnamefont
  {Lucignano}},\ }\bibfield  {title} {\bibinfo {title} {Quantum transport in
  rashba spin{\textendash}orbit materials: a review},\ }\href
  {https://doi.org/10.1088/0034-4885/78/10/106001} {\bibfield  {journal}
  {\bibinfo  {journal} {Reports on Progress in Physics}\ }\textbf {\bibinfo
  {volume} {78}},\ \bibinfo {pages} {106001} (\bibinfo {year}
  {2015})}\BibitemShut {NoStop}%
\bibitem [{\citenamefont {Manchon}\ \emph {et~al.}(2015)\citenamefont
  {Manchon}, \citenamefont {Koo}, \citenamefont {Nitta}, \citenamefont
  {Frolov},\ and\ \citenamefont {Duine}}]{rashba_review}%
  \BibitemOpen
  \bibfield  {author} {\bibinfo {author} {\bibfnamefont {A.}~\bibnamefont
  {Manchon}}, \bibinfo {author} {\bibfnamefont {H.~C.}\ \bibnamefont {Koo}},
  \bibinfo {author} {\bibfnamefont {J.}~\bibnamefont {Nitta}}, \bibinfo
  {author} {\bibfnamefont {S.~M.}\ \bibnamefont {Frolov}},\ and\ \bibinfo
  {author} {\bibfnamefont {R.~A.}\ \bibnamefont {Duine}},\ }\bibfield  {title}
  {\bibinfo {title} {{N}ew perspectives for {R}ashba spin–orbit coupling},\
  }\href@noop {} {\bibfield  {journal} {\bibinfo  {journal} {Nature materials}\
  }\textbf {\bibinfo {volume} {14}},\ \bibinfo {pages} {871} (\bibinfo {year}
  {2015})}\BibitemShut {NoStop}%
\bibitem [{\citenamefont {Riera}(2013)}]{rashbahub_PhysRevB.88.045102}%
  \BibitemOpen
  \bibfield  {author} {\bibinfo {author} {\bibfnamefont {J.~A.}\ \bibnamefont
  {Riera}},\ }\bibfield  {title} {\bibinfo {title} {Spin polarization in the
  hubbard model with rashba spin-orbit coupling on a ladder},\ }\href
  {https://doi.org/10.1103/PhysRevB.88.045102} {\bibfield  {journal} {\bibinfo
  {journal} {Phys. Rev. B}\ }\textbf {\bibinfo {volume} {88}},\ \bibinfo
  {pages} {045102} (\bibinfo {year} {2013})}\BibitemShut {NoStop}%
\bibitem [{\citenamefont {Nolting}\ and\ \citenamefont
  {Brewer}(2009)}]{book_Nolting}%
  \BibitemOpen
  \bibfield  {author} {\bibinfo {author} {\bibfnamefont {W.}~\bibnamefont
  {Nolting}}\ and\ \bibinfo {author} {\bibfnamefont {W.}~\bibnamefont
  {Brewer}},\ }\href@noop {} {\emph {\bibinfo {title} {Fundamentals of
  Many-body Physics: Principles and Methods}}}\ (\bibinfo  {publisher}
  {Springer Berlin Heidelberg},\ \bibinfo {year} {2009})\BibitemShut {NoStop}%
\bibitem [{\citenamefont {Zhang}\ \emph {et~al.}(2015)\citenamefont {Zhang},
  \citenamefont {Jungfleisch}, \citenamefont {Jiang}, \citenamefont {Liu},
  \citenamefont {Pearson}, \citenamefont {Velthuis}, \citenamefont {Hoffmann},
  \citenamefont {Freimuth},\ and\ \citenamefont
  {Mokrousov}}]{PhysRevB.91.115316}%
  \BibitemOpen
  \bibfield  {author} {\bibinfo {author} {\bibfnamefont {W.}~\bibnamefont
  {Zhang}}, \bibinfo {author} {\bibfnamefont {M.~B.}\ \bibnamefont
  {Jungfleisch}}, \bibinfo {author} {\bibfnamefont {W.}~\bibnamefont {Jiang}},
  \bibinfo {author} {\bibfnamefont {Y.}~\bibnamefont {Liu}}, \bibinfo {author}
  {\bibfnamefont {J.~E.}\ \bibnamefont {Pearson}}, \bibinfo {author}
  {\bibfnamefont {S.~G. E.~t.}\ \bibnamefont {Velthuis}}, \bibinfo {author}
  {\bibfnamefont {A.}~\bibnamefont {Hoffmann}}, \bibinfo {author}
  {\bibfnamefont {F.}~\bibnamefont {Freimuth}},\ and\ \bibinfo {author}
  {\bibfnamefont {Y.}~\bibnamefont {Mokrousov}},\ }\bibfield  {title} {\bibinfo
  {title} {Reduced spin-hall effects from magnetic proximity},\ }\href
  {https://doi.org/10.1103/PhysRevB.91.115316} {\bibfield  {journal} {\bibinfo
  {journal} {Phys. Rev. B}\ }\textbf {\bibinfo {volume} {91}},\ \bibinfo
  {pages} {115316} (\bibinfo {year} {2015})}\BibitemShut {NoStop}%
\bibitem [{\citenamefont {Zhang}\ \emph {et~al.}(2011)\citenamefont {Zhang},
  \citenamefont {Freimuth}, \citenamefont {Bl\"ugel}, \citenamefont
  {Mokrousov},\ and\ \citenamefont {Souza}}]{PhysRevLett.106.117202}%
  \BibitemOpen
  \bibfield  {author} {\bibinfo {author} {\bibfnamefont {H.}~\bibnamefont
  {Zhang}}, \bibinfo {author} {\bibfnamefont {F.}~\bibnamefont {Freimuth}},
  \bibinfo {author} {\bibfnamefont {S.}~\bibnamefont {Bl\"ugel}}, \bibinfo
  {author} {\bibfnamefont {Y.}~\bibnamefont {Mokrousov}},\ and\ \bibinfo
  {author} {\bibfnamefont {I.}~\bibnamefont {Souza}},\ }\bibfield  {title}
  {\bibinfo {title} {Role of spin-flip transitions in the anomalous hall effect
  of fept alloy},\ }\href {https://doi.org/10.1103/PhysRevLett.106.117202}
  {\bibfield  {journal} {\bibinfo  {journal} {Phys. Rev. Lett.}\ }\textbf
  {\bibinfo {volume} {106}},\ \bibinfo {pages} {117202} (\bibinfo {year}
  {2011})}\BibitemShut {NoStop}%
\bibitem [{\citenamefont {Mandal}\ \emph {et~al.}(2019)\citenamefont {Mandal},
  \citenamefont {Haule}, \citenamefont {Rabe},\ and\ \citenamefont
  {Vanderbilt}}]{beyond_dft}%
  \BibitemOpen
  \bibfield  {author} {\bibinfo {author} {\bibfnamefont {S.}~\bibnamefont
  {Mandal}}, \bibinfo {author} {\bibfnamefont {K.}~\bibnamefont {Haule}},
  \bibinfo {author} {\bibfnamefont {K.~M.}\ \bibnamefont {Rabe}},\ and\
  \bibinfo {author} {\bibfnamefont {D.}~\bibnamefont {Vanderbilt}},\ }\bibfield
   {title} {\bibinfo {title} {Systematic beyond-dft study of binary transition
  metal oxides},\ }\href@noop {} {\bibfield  {journal} {\bibinfo  {journal}
  {npj Computational Materials}\ }\textbf {\bibinfo {volume} {5}},\ \bibinfo
  {pages} {115} (\bibinfo {year} {2019})}\BibitemShut {NoStop}%
\bibitem [{\citenamefont {Shick}\ \emph {et~al.}(1999)\citenamefont {Shick},
  \citenamefont {Liechtenstein},\ and\ \citenamefont
  {Pickett}}]{PhysRevB.60.10763}%
  \BibitemOpen
  \bibfield  {author} {\bibinfo {author} {\bibfnamefont {A.~B.}\ \bibnamefont
  {Shick}}, \bibinfo {author} {\bibfnamefont {A.~I.}\ \bibnamefont
  {Liechtenstein}},\ and\ \bibinfo {author} {\bibfnamefont {W.~E.}\
  \bibnamefont {Pickett}},\ }\bibfield  {title} {\bibinfo {title}
  {Implementation of the lda+u method using the full-potential linearized
  augmented plane-wave basis},\ }\href
  {https://doi.org/10.1103/PhysRevB.60.10763} {\bibfield  {journal} {\bibinfo
  {journal} {Phys. Rev. B}\ }\textbf {\bibinfo {volume} {60}},\ \bibinfo
  {pages} {10763} (\bibinfo {year} {1999})}\BibitemShut {NoStop}%
\bibitem [{\citenamefont {M\l{}y\ifmmode~\acute{n}\else \'{n}\fi{}czak}\ \emph
  {et~al.}(2022)\citenamefont {M\l{}y\ifmmode~\acute{n}\else \'{n}\fi{}czak},
  \citenamefont {Aguilera}, \citenamefont {Gospodari\ifmmode~\check{c}\else
  \v{c}\fi{}}, \citenamefont {Heider}, \citenamefont {Jugovac}, \citenamefont
  {Zamborlini}, \citenamefont {Hanke}, \citenamefont {Friedrich}, \citenamefont
  {Mokrousov}, \citenamefont {Tusche}, \citenamefont {Suga}, \citenamefont
  {Feyer}, \citenamefont {Bl\"ugel}, \citenamefont {Plucinski},\ and\
  \citenamefont {Schneider}}]{PhysRevB.105.115135}%
  \BibitemOpen
  \bibfield  {author} {\bibinfo {author} {\bibfnamefont {E.}~\bibnamefont
  {M\l{}y\ifmmode~\acute{n}\else \'{n}\fi{}czak}}, \bibinfo {author}
  {\bibfnamefont {I.}~\bibnamefont {Aguilera}}, \bibinfo {author}
  {\bibfnamefont {P.}~\bibnamefont {Gospodari\ifmmode~\check{c}\else
  \v{c}\fi{}}}, \bibinfo {author} {\bibfnamefont {T.}~\bibnamefont {Heider}},
  \bibinfo {author} {\bibfnamefont {M.}~\bibnamefont {Jugovac}}, \bibinfo
  {author} {\bibfnamefont {G.}~\bibnamefont {Zamborlini}}, \bibinfo {author}
  {\bibfnamefont {J.-P.}\ \bibnamefont {Hanke}}, \bibinfo {author}
  {\bibfnamefont {C.}~\bibnamefont {Friedrich}}, \bibinfo {author}
  {\bibfnamefont {Y.}~\bibnamefont {Mokrousov}}, \bibinfo {author}
  {\bibfnamefont {C.}~\bibnamefont {Tusche}}, \bibinfo {author} {\bibfnamefont
  {S.}~\bibnamefont {Suga}}, \bibinfo {author} {\bibfnamefont {V.}~\bibnamefont
  {Feyer}}, \bibinfo {author} {\bibfnamefont {S.}~\bibnamefont {Bl\"ugel}},
  \bibinfo {author} {\bibfnamefont {L.}~\bibnamefont {Plucinski}},\ and\
  \bibinfo {author} {\bibfnamefont {C.~M.}\ \bibnamefont {Schneider}},\
  }\bibfield  {title} {\bibinfo {title} {Fe(001) angle-resolved photoemission
  and intrinsic anomalous hall conductivity in fe seen by different ab initio
  approaches: Lda and gga versus $\mathit{GW}$},\ }\href
  {https://doi.org/10.1103/PhysRevB.105.115135} {\bibfield  {journal} {\bibinfo
   {journal} {Phys. Rev. B}\ }\textbf {\bibinfo {volume} {105}},\ \bibinfo
  {pages} {115135} (\bibinfo {year} {2022})}\BibitemShut {NoStop}%
\end{thebibliography}%

\end{document}